%% file: manuscript_v3.tex
\documentclass[aps,pra,twocolumn,superscriptaddress,floatfix,preprintnumbers,10pt]{revtex4-2}

\usepackage[T1]{fontenc}
\usepackage{amsmath,amssymb,amsfonts,mathtools,bm,graphicx}
\usepackage[colorlinks=true,linkcolor=blue,citecolor=magenta,urlcolor=blue]{hyperref}
\usepackage{orcidlink}

\hypersetup{
  pdftitle={Emergent hydrodynamic response and dynamical backreaction: Magnon bound-state propagation in a Bose-Hubbard fluid},
  pdfauthor={A. N. Caliz, A. Riera, E. Rico, J. Barata, M. Plodzien}}

\graphicspath{{figures/}}

\newcommand{\ii}{\mathrm{i}}
\newcommand{\dd}{\mathrm{d}}
\newcommand{\hc}{\mathrm{H.c.}}
\newcommand{\Sp}{\hat S^{+}}
\newcommand{\Sm}{\hat S^{-}}
\newcommand{\Sz}{\hat S^{z}}
\newcommand{\ad}{\hat a^{\dagger}}
\newcommand{\nm}{\hat m}
\newcommand{\nbar}{\bar n}
\newcommand{\HXXZ}{\hat H_{\mathrm{XXZ}}}
\newcommand{\HBH}{\hat H_{\mathrm{BH}}}
\newcommand{\HSB}{\hat H_{\mathrm{SB}}}
\newcommand{\avg}[1]{\left\langle #1\right\rangle}
\newcommand{\half}{\tfrac12}
\newcommand{\Imm}{\operatorname{Im}}
\newcommand{\vsc}{v_{\mathrm{sc}}}

\providecommand{\ket}[1]{\left|#1\right\rangle}

\begin{document}

\title{Emergent hydrodynamic response and dynamical backreaction:\\
Magnon bound-state propagation in a Bose--Hubbard fluid}

\author{Andrés N. Cáliz\,\orcidlink{0009-0002-6938-8595}}
\email{andres.navas@qilimanjaro.tech}
\affiliation{Qilimanjaro Quantum Tech, Carrer de Vene\c{c}uela 74, 08019 Barcelona, Spain}
\affiliation{Departament de Física, Universitat de Barcelona, 08007 Barcelona, Spain}

\author{Arnau Riera\,\orcidlink{0000-0002-3271-7802}}
\affiliation{Qilimanjaro Quantum Tech, Carrer de Vene\c{c}uela 74, 08019 Barcelona, Spain}

\author{Enrique Rico\,\orcidlink{0000-0003-4414-6821}}
\affiliation{Theoretical Physics Department, CERN, 1211 Geneva 23, Switzerland}

\author{Jo\~{a}o Barata\,\orcidlink{0000-0003-4286-4555}}
\affiliation{Theoretical Physics Department, CERN, 1211 Geneva 23, Switzerland}

\author{Marcin P\l odzie\'n\,\orcidlink{0000-0002-0835-1644}}
\email{marcin.plodzien@qilimanjaro.tech}
\affiliation{Qilimanjaro Quantum Tech, Carrer de Vene\c{c}uela 74, 08019 Barcelona, Spain}

\date{\today}

\preprint{CERN-TH-2026-192}

\input{sec_v3/00_abstract}

\maketitle

\input{sec_v3/01_introduction}
\input{sec_v3/02_model}
\input{sec_v3/03_hydrodynamics}
\input{sec_v3/04_numerics}
\input{sec_v3/05_discussion}
\input{sec_v3/06_acknowledgments}

\bibliographystyle{apsrev4-2}
\bibliography{refs_v3}

\appendix
\onecolumngrid

\input{sec_v3/appendix_bose_hubbard}
\input{sec_v3/appendix_spin_boson}
\input{sec_v3/appendix_response}
\input{sec_v3/appendix_two_magnon}
\input{sec_v3/appendix_numerics}
\end{document}

%% file: sec_v3/00_abstract.tex
\begin{abstract}
We study the nonequilibrium response of a one-dimensional Bose--Hubbard medium to a two-magnon bound state propagating along an attractive XXZ chain. A Holstein-type displacement coupling makes the magnon density act both as a local chemical-potential perturbation and as a source of bosons. We derive a long-wavelength hydrodynamic description of the density and phase fluctuations and test it against matrix-product-state simulations of the full coupled dynamics. The theory predicts a comoving near-field deformation together with retarded density waves confined to a sound cone. In the Mott regime, the deformation remains localized around the moving pair. In the compressible regime, two counterpropagating fronts detach and approach the semiclassical sound velocity, recovered from the equilibrium density with no fitted parameters. The coupling also induces a dynamical backreaction that slows the bound state and broadens its magnon-density profile. The comparison delimits where hydrodynamics stays quantitative once the probe reacts back on the medium.
\end{abstract}

%% file: sec_v3/01_introduction.tex
\section{Introduction}\label{sec:intro}

A mobile quasiparticle coupled to a compressible quantum fluid can generate two qualitatively distinct responses. It produces a deformation that remains attached to it, forming the dressing cloud underlying polaron formation~\cite{Bruderer2007,CastroNeto1996,Lamacraft2009}, and may also emit collective excitations that propagate independently through the medium~\cite{Astrakharchik2004,Carusotto2006}. In one spatial dimension, the long-wavelength dynamics of these detached excitations are governed by Luttinger liquid hydrodynamics~\cite{Haldane1981,Giamarchi2003,Cazalilla2011,Imambekov2012}, and their leading density fronts propagate at the sound velocity of the fluid. Experiments with impurities in Bose gases have made this interplay accessible~\cite{Catani2012,Meinert2017,Jorgensen2016}. The Bose--Hubbard chain provides a complementary lattice setting in which the medium can be tuned from an incompressible Mott insulator to a compressible superfluid~\cite{Fisher1989,Jaksch1998,Greiner2002}.

A characteristic signature of the hydrodynamic response is the formation of a sound cone in the spacetime density profile~\cite{Cheneau2012,Carleo2014}. In a compressible medium, density disturbances detach from a moving quasiparticle and form counterpropagating fronts, signaling collective modes that propagate independently of the source. This behavior can be characterized using a long-wavelength density-phase description of the bosonic sector, from which one can extract quantities such as the retarded response, compressibility, phase stiffness, sound velocity, and effective source couplings. Going beyond this long-wavelength description generally requires solving the real-time quantum dynamics of the coupled probe-medium system, thereby allowing one to test the emergence and range of validity of the hydrodynamic description.

In this work, we compare the predictions of an effective hydrodynamic
description with microscopic quantum many-body simulations based on matrix product states (MPSs)~\cite{White1992,Schollwock2011,Fishman2022,Haegeman2011,Haegeman2016,Paeckel2019}. We use a fully dynamical probe, namely a composite quasiparticle hosted by an interacting spin chain, rather than an externally imposed potential following a prescribed trajectory. Specifically, we consider an attractive XXZ chain in which two magnons form the simplest mobile bound state in the model~\cite{Bethe1931,Wortis1963,Ganahl2012,Fukuhara2013}. Its internal profile and dispersion can both be obtained analytically. Mobile spin excitations of this kind have also been resolved in quantum-gas microscopes~\cite{FukuharaMobile2013}. We prepare the bound pair as a localized wave packet with finite center-of-mass momentum and release it into the coupled system, thereby initiating a unitary nonequilibrium evolution. In previous work, we studied the complementary case of an XX chain, in which the Bose--Hubbard bath dresses magnon motion and mediates an attraction between otherwise unbound magnons~\cite{Caliz2026Dressed}. Here, by contrast, an intrinsically bound pair in an attractive XXZ chain probes the medium's hydrodynamic response and the resulting backreaction across the Mott--superfluid transition.

The bound pair is coupled locally to the displacement of the Bose--Hubbard field through an interaction analogous to the Holstein coupling~\cite{Holstein1959,Plodzien2018,Kosior2023,Plodzien2026Sensitivity}. In the compressible regime, the infrared dynamics of the Bose--Hubbard model are described by Luttinger liquid theory~\cite{Haldane1981,MattisLieb1965,Luttinger1963,Tomonaga1950}. Within this description, the interaction generates two distinct source channels. It shifts the local energy required to add a boson, thereby acting as a position- and time-dependent chemical potential, and introduces an explicit source term into the local boson-number continuity equation. The motion of the pair further separates the response into a near-field deformation that follows the magnons and density disturbances that detach from the source and propagate through the medium. Because the bound state remains a dynamical degree of freedom, the interaction
also modifies its velocity and spatial density profile. We refer to these changes as the backreaction of the medium on the probe. Capturing this backreaction fully requires going beyond the linearized hydrodynamic treatment of the medium response.

For the Bose--Hubbard medium, we find a clear change in the response across the Mott--superfluid transition. In the Mott regime, the induced density deformation remains localized around the moving pair because the background lacks a gapless acoustic channel. In the compressible regime, two counterpropagating fronts detach from the source and travel at velocities approaching the semiclassical sound velocity, while a narrower deformation remains attached to the magnons. As the spin-boson coupling increases, the fluid exerts a stronger backreaction on the bound state, reducing its propagation velocity and broadening its magnon-density profile.

The paper is organized as follows. Section~\ref{sec:model} introduces the lattice model, the two-magnon bound state, and the preparation protocol. Section~\ref{sec:hydro} derives the effective hydrodynamic theory and the corresponding response, while Section~\ref{sec:num} compares these predictions with the results of the many-body simulations. Finally, Section~\ref{sec:disc} presents our conclusions. Technical derivations and details of the tensor-network calculations are provided in the appendices.

%% file: sec_v3/02_model.tex
\section{Magnon bound state in a Bose--Hubbard medium}\label{sec:model}

We consider an open chain of $L$ composite sites, each carrying a spin-$1/2$ degree of freedom and a bosonic mode. With $\hbar=1$, the dynamics of the coupled system are governed by the Hamiltonian
$\hat H=\HXXZ+\HBH+\HSB$, where
\begin{subequations}\label{eq:H}
\begin{align}
  \HXXZ
  &=\frac J2\sum_{i=1}^{L-1}
  \big(\Sp_i\Sm_{i+1}+\hc\big)
  \nonumber\\
  &\quad
  +J\Delta\sum_{i=1}^{L-1}\Sz_i\Sz_{i+1}
  +Jh_z\sum_{i=1}^{L}\Sz_i,
  \label{eq:HXXZ}\\
  \HBH
  &=-t_B\sum_{i=1}^{L-1}
  \big(\ad_i\hat a_{i+1}+\hc\big)
  \nonumber\\
  &\quad
  +\frac{U_B}{2}\sum_{i=1}^{L}
  \hat n_i(\hat n_i-1)
  -\mu\sum_{i=1}^{L}\hat n_i,
  \label{eq:HBH}\\
  \HSB
  &=\lambda_z\sum_{i=1}^{L}
  \nm_i\big(\hat a_i+\ad_i\big).
  \label{eq:HSB}
\end{align}
\end{subequations}
Here, $\Sp_i$, $\Sm_i$, and $\Sz_i$ are spin-$1/2$ operators, $\hat a_i$ annihilates a boson on site $i$, and $\hat n_i=\ad_i\hat a_i$ is the local boson-number operator. 
\newline
The local magnon density is defined by
\begin{equation}\label{eq:magnon_density}
  \nm_i=\Sz_i+\half \, .
\end{equation}
The magnon density is measured relative to the fully down-polarized spin vacuum and takes the value one on a flipped site and zero elsewhere. Consequently, the magnon-free spin chain does not drive the bosonic medium. Although the coupling strength $\lambda_z$ is spatially uniform, the interaction acts only where the magnon density is nonzero and therefore constitutes a localized drive. Coupling directly to $\Sz_i$ would differ from Eq.~\eqref{eq:HSB} by a spatially uniform displacement term and would shift the equilibrium configuration of the bosonic medium. The definition in Eq.~\eqref{eq:magnon_density} avoids this uniform drive in the fully polarized vacuum. Since $[\hat H,\sum_i\nm_i]=0$, the magnon number is conserved, and we work throughout in the sector $\sum_i\nm_i=2$. By contrast, the boson number is not conserved because
$[\hat H,\sum_i\hat n_i]\neq0$. This nonconservation follows from the linear dependence of $\HSB$ on $\hat a_i+\ad_i$ and gives rise to the source terms appearing in the hydrodynamic continuity equation discussed in Section~\ref{sec:hydro} and in the exact local boson-number balance law derived in Appendix~\ref{app:sb} and tested in Section~\ref{sec:num}. The bosonic sector can be parametrized by
\begin{equation}\label{eq:ratios}
  r=\frac{t_B}{U_B},
  \qquad
  s=\frac{\mu}{U_B},
\end{equation}
together with the dimensionless spin-boson coupling
\begin{equation}\label{eq:eta}
  \eta=\frac{\lambda_z}{U_B}.
\end{equation}
We measure energies in units of $U_B$ and times in units of $U_B^{-1}$, and set $h_z=0$ throughout this work. At integer filling and small $r$, the isolated Bose--Hubbard chain is an incompressible Mott insulator, whereas outside the Mott lobes its ground state is a compressible Luttinger liquid~\cite{Fisher1989,Kuhner1998,Ejima2012,Kollath2005}. Below we follow the line $s=0.5$, which crosses from the unit-filling Mott lobe into the superfluid regime. The location of this crossing is determined independently using the equilibrium analysis in Appendix~\ref{app:bh}. For $J>0$, the spin chain lies in the attractive regime $\Delta<0$, in which two neighboring magnons are lower in energy than two separated magnons by $|J\Delta|$ and form a bound state.

\subsection{Two-magnon bound state}
In the two-magnon sector, a basis state can be written as
\begin{equation}\label{eq:basis}
  \ket{x_1,x_2}
  =\Sp_{x_1}\Sp_{x_2}\ket{\Downarrow},
  \qquad x_1<x_2,
\end{equation}
where $\ket{\Downarrow}$ denotes the fully down-polarized vacuum. Within the coordinate Bethe ansatz, the bound-state branch is described by a pair of complex magnon quasimomenta whose imaginary parts localize the relative coordinate~\cite{Bethe1931,Wortis1963}. Defining the center-of-mass and relative coordinates as
\begin{equation}
  X=\frac{x_1+x_2}{2},
  \qquad
  \ell=x_2-x_1,
\end{equation}
the bound state at total momentum $K$ has the exponentially localized profile
\begin{equation}\label{eq:profile}
  \psi_K(x_1,x_2)\propto
  e^{\ii KX}A_K^{\ell-1},
  \qquad
  A_K=\frac{\cos(K/2)}{\Delta}.
\end{equation}
The state is normalizable when $|A_K|<1$. Its dispersion and group velocity are
\begin{subequations}\label{eq:dispersion}
\begin{align}
  E_{\mathrm{B}}(K)
  &=J(2h_z-\Delta)
  +\frac{J}{\Delta}\cos^2\frac K2,
  \label{eq:dispersion_E}\\
  v_{\mathrm{B}}(K)
  &=-\frac{J\sin K}{2\Delta}.
  \label{eq:dispersion_v}
\end{align}
\end{subequations}
The subscript B on $E_{\mathrm{B}}$ and $v_{\mathrm{B}}$ labels the bound state, whereas on
$t_B$, $U_B$, and $\HBH$ it labels the bosonic medium.
For $\Delta<0$ and $0<K<\pi$, the group velocity is positive and can be written as $v_{\mathrm{B}}(K)=J\sin K/(2|\Delta|)$. At $K_0=\pi/2$, the curvature $E_{\mathrm{B}}''(K_0)$ vanishes, suppressing the leading dispersive spreading of a wave packet narrow in momentum space. On an infinite chain, the mean separation between the two magnons is
\begin{equation}\label{eq:mean_separation}
  \avg{\ell}_K
  =\frac{1}{1-|A_K|^2}.
\end{equation}
This quantity defines the characteristic internal length of the ideal bound state. It should be distinguished from the spatial width of a localized magnon-density packet, which also contains the center-of-mass envelope and is the quantity analyzed in Section~\ref{sec:num}. The derivation of Eqs.~\eqref{eq:profile}--\eqref{eq:mean_separation} is given in Appendix~\ref{app:two_magnon}.

\subsection{State preparation and evolution}

\begin{figure}[t]
  \centering
  \includegraphics[width=\columnwidth]{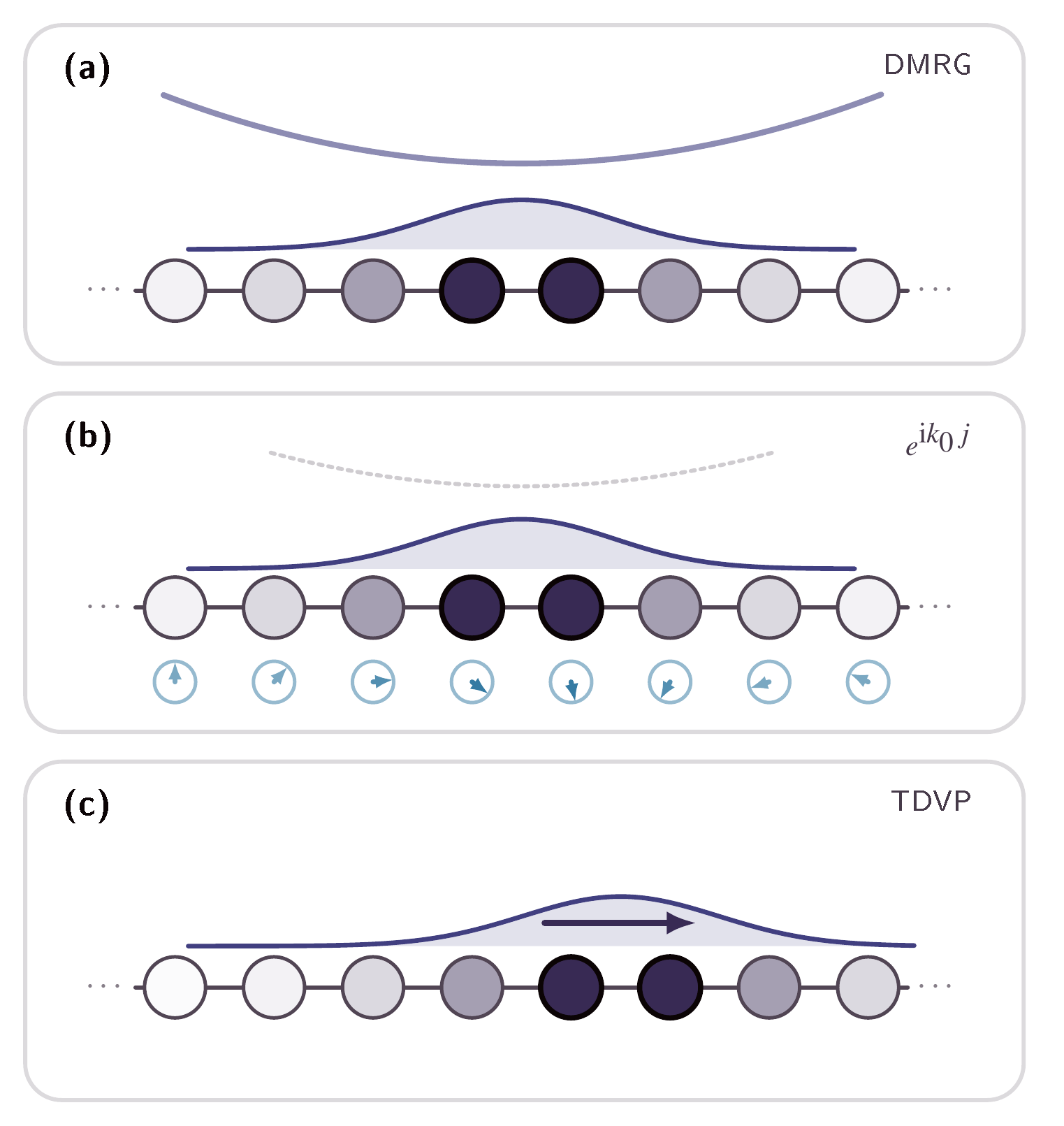}
  \caption{Tensor-network preparation and evolution protocol. (a) The density-matrix renormalization group (DMRG) prepares a trapped two-magnon bound state in the sector $\sum_i\nm_i=2$. The harmonic trap confines its center-of-mass envelope, while $\Delta_{\mathrm{prep}}$ controls its relative profile. (b) The trap is removed, and the diagonal kick in Eq.~\eqref{eq:kick} imprints the momentum $K_0=2k_0$. The matching condition in Eq.~\eqref{eq:matching} minimizes the resulting internal excitation. (c) The spin-boson coupling is switched on, and the time-dependent variational principle (TDVP) evolves the state under the Hamiltonian in Eq.~\eqref{eq:H}. Numerical details are given in Appendix~\ref{app:num}.}
  \label{fig:tn}
\end{figure}

To prepare the initial condition for the evolution of the system, we employ a trap-and-kick protocol summarized in Fig.~\ref{fig:tn}. In the absence of the spin-boson coupling, we first prepare a stationary two-magnon bound state in a weak harmonic trap centered at $X_0$. The trap localizes the center-of-mass envelope while leaving the relative profile close to that of the $K=0$ bound state. A finite center-of-mass momentum is subsequently imprinted through the diagonal unitary transformation
\begin{equation}\label{eq:kick}
  \hat U_{\mathrm{kick}}(k_0)
  =\prod_{j=1}^{L}
  \exp\!\big(\ii k_0 j\,\nm_j\big).
\end{equation}
Acting on the basis state in Eq.~\eqref{eq:basis}, the kick produces a phase $e^{\ii k_0(x_1+x_2)}=e^{\ii(2k_0)X}$. Each magnon therefore acquires momentum $k_0$, and the total momentum of the pair is $K_0=2k_0$. As the kick is diagonal in the magnon positions, it changes only the center-of-mass phase and leaves the relative profile unchanged. Matching the relative profile centered at $K=0$ during preparation to the profile selected by Eq.~\eqref{eq:profile} at momentum $K_0$ requires
\begin{equation}\label{eq:matching}
  \Delta_{\mathrm{prep}}
  =
  \frac{\Delta_{\mathrm{evol}}}{\cos(K_0/2)}.
\end{equation}
For $K_0=\pi/2$, this yields $\Delta_{\mathrm{prep}}=\sqrt{2}\,\Delta_{\mathrm{evol}}$, with both anisotropies negative. Since the trapped state has a finite momentum width, Eq.~\eqref{eq:matching} matches the relative profile only at the central momentum of the wave packet. It therefore minimizes the internal quench induced by the anisotropy change. Subsequent changes in the spatial magnon-density profile can then be interpreted in terms of the finite momentum width, higher-order lattice dispersion, and coupling to the bosonic medium.

The initial state of the composite chain is obtained in a single DMRG calculation for the decoupled preparation Hamiltonian and is therefore equivalent to the product of the trapped two-magnon state and the Bose--Hubbard ground state at the same $(r,s)$. At $t=0$, the trap is removed, the momentum kick is applied, and the spin-boson coupling is quenched from $\lambda_z=0$ to its target value. The subsequent dynamics are governed by the full Hamiltonian in Eq.~\eqref{eq:H} and computed using TDVP~\cite{Haegeman2011,Haegeman2016,Paeckel2019} in the ITensor framework~\cite{Fishman2022}. This constitutes a global quench of the coupling strength but produces a spatially localized drive because $\nm_i$ is nonzero only around the bound pair. Each MPS site combines the spin-$1/2$ degree of freedom with a truncated bosonic Fock space. Magnon-number conservation is implemented as a global $U(1)$ symmetry, whereas the nonconserved boson number cannot be used as a quantum number because of $\HSB$. Further implementation and convergence details are provided in Appendix~\ref{app:num}.

%% file: sec_v3/03_hydrodynamics.tex
\section{Hydrodynamic theory of the bosonic response}\label{sec:hydro}

In this section, we derive the long-wavelength response of the Bose--Hubbard fluid to the moving two-magnon bound state. Starting from Eq.~\eqref{eq:H}, we relate the effective parameters to the microscopic couplings and separate the quasistatic deformation from the propagating response generated by switching on the coupling and by the subsequent motion of the pair. Details are given in Appendices~\ref{app:bh}, \ref{app:sb}, and \ref{app:response}.

\subsection{Density-phase reduction}\label{sec:hydro_density}

We start by writing the lattice bosons in rotor variables,
\begin{equation}\label{eq:rotor}
  \hat n_i=\nbar+\hat\Pi_i,
  \qquad
  \hat a_i\simeq e^{-\ii\hat\theta_i}\sqrt{\nbar+\hat\Pi_i},
\end{equation}
with $[\hat\theta_i,\hat\Pi_j]=\ii\delta_{ij}$ and, in the continuum,
$[\hat\theta(x),\hat\rho(y)]=\ii\delta(x-y)$, where
$\hat\Pi_i=a\hat\rho(x_i)$. With this convention, the current is
$\hat j=-\rho_s\partial_x\hat\theta$. Local amplitude-phase products are
understood in the symmetrized semiclassical ordering of the rotor expansion. Expanding Eq.~\eqref{eq:HBH} to quadratic order and canceling the terms linear in $\hat\Pi_i$ in the bulk gives
\begin{equation}\label{eq:Hlat}
  \hat H_{\mathrm{B}}^{\mathrm{IR}}
  =\frac{U_B}{2}\sum_{i=1}^{L}\hat\Pi_i^2
  +t_B\nbar\sum_{i=1}^{L-1}
  \big(\hat\theta_{i+1}-\hat\theta_i\big)^2 .
\end{equation}
Here, we omit a constant and a density-gradient term
$\propto(\hat\Pi_{i+1}-\hat\Pi_i)^2$, which is subleading in the infrared. Using $\sum_i\to a^{-1}\int\dd x$ and $\hat\theta_{i+1}-\hat\theta_i\to a\partial_x\hat\theta$ yields
\begin{equation}\label{eq:Hhyd}
  \hat H_{\mathrm{B}}^{\mathrm{hyd}}
  =\frac12\int\dd x
  \left[
    \chi^{-1}\hat\rho^2
    +\rho_s(\partial_x\hat\theta)^2
  \right],
\end{equation}
with 
\begin{equation}\label{eq:chirhos}
  \chi^{-1}=U_Ba,
  \qquad
  \rho_s=2t_B\nbar a .
\end{equation}
Therefore,
\begin{align}\label{eq:vsound}
  &\vsc=a\sqrt{2U_Bt_B\nbar},
  \qquad
  \frac{\vsc}{aU_B}=\sqrt{2r\nbar}\, , \\ 
  \notag 
  &\mathcal K_{\mathrm{sc}}
  =\pi\sqrt{\rho_s\chi}
  =\pi\sqrt{2r\nbar}\, .
\end{align}
Here, we use $\mathcal K$ for the Luttinger parameter and $K$ for the pair momentum. Equations~\eqref{eq:chirhos}--\eqref{eq:vsound} are semiclassical estimates that provide a valid description of the low-energy dynamics of the theory. The Luttinger liquid form in Eq.~\eqref{eq:Hhyd} survives in the compressible phase, but its coefficients are renormalized, particularly near the Mott transition~\cite{Haldane1981,Giamarchi2003,Kuhner1998,Ejima2012}, leading to different quantitative predictions. In what follows, we test Eq.~\eqref{eq:vsound} using the equilibrium density $\nbar(r,s)$ directly.

\subsection{The magnon density as a source}\label{sec:hydro_source}

\begin{figure}[h!]
  \centering
  \includegraphics[width=\columnwidth]{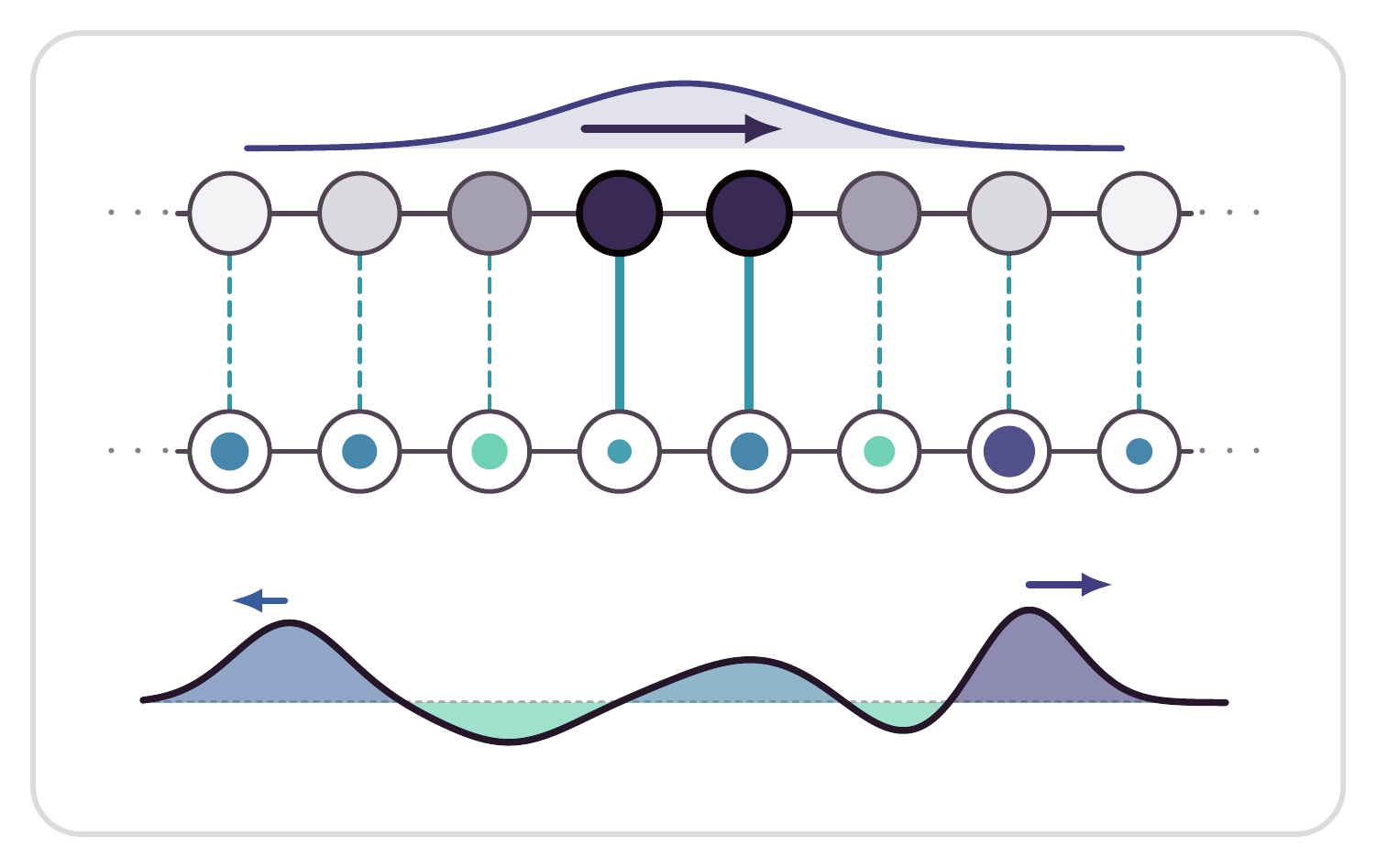}
  \caption{Geometry of the coupled system and its hydrodynamic response. The
  displacement coupling acts only where the two-magnon density is nonzero. The
  induced density contains a near-field deformation that follows the pair and
  counterpropagating fronts generated by the switch-on and subsequent motion.
  Their leading edges propagate at the sound velocity.}
  \label{fig:diagram}
\end{figure}

Within this formulation, we can characterize the medium's response to the magnon source. To this end, we write $\hat\theta_i=\theta_0+\hat\vartheta_i$ and expand the displacement operator to first order:
\begin{equation}\label{eq:dispexp}
  \hat a_i+\ad_i
  \simeq
  2\sqrt{\nbar}\cos\theta_0
  -2\sqrt{\nbar}\sin\theta_0\,\hat\vartheta_i
  +\frac{\cos\theta_0}{\sqrt{\nbar}}\hat\Pi_i .
\end{equation}
The first term is constant at fixed magnon number. Introducing the continuum magnon density
\begin{equation}\label{eq:mcont}
  \hat m(x)=\sum_{i=1}^{L}\nm_i\delta(x-x_i),
\end{equation}
the remaining terms define
\begin{equation}\label{eq:Hsrc}
  \hat H_{\mathrm{src}}
  =g_\theta\int\dd x\,\hat m\hat\vartheta
  +ag_\rho\int\dd x\,\hat m\hat\rho ,
\end{equation}
where $g_\theta=-2\lambda_z\sqrt{\nbar}\sin\theta_0$ and $g_\rho=\frac{\lambda_z}{\sqrt{\nbar}}\cos\theta_0$. Here, $g_\rho$ generates a local chemical-potential shift, whereas $g_\theta$ is a
number-changing source in the continuity equation. Dropping the constant energy shift, we obtain the complete reduced Hamiltonian
\begin{equation}\label{eq:HIR}
\begin{aligned}
  \hat H_{\mathrm{IR}}
  &=
  \frac12\int\dd x
  \left[
    \chi^{-1}\hat\rho^2
    +\rho_s(\partial_x\hat\vartheta)^2
  \right]
  \\
  &\quad
  +g_\theta\int\dd x\,\hat m\hat\vartheta
  +ag_\rho\int\dd x\,\hat m\hat\rho \, ,
\end{aligned}
\end{equation}
containing both the intrinsic hydrodynamics of the fluid and the
two source channels generated by the moving magnon density. For the fixed-phase semiclassical sector, we choose $\theta_0=0$, giving
$g_\theta=0$ and $g_\rho=\lambda_z/\sqrt{\nbar}$. The fixed-number
Bose--Hubbard state used in DMRG has no definite global phase. The fixed-phase
theory therefore describes a semiclassical sector, and global $U(1)$ symmetry is
represented in the nonlinear initial condition by phase averaging. The
microscopic coupling itself remains number-changing: the first surviving phase
term is $-\lambda_z\sqrt{\nbar}\sum_i\nm_i\hat\vartheta_i^2$ and contributes
from $O(\lambda_z^2)$ onward; see Appendix~\ref{app:sb}. Using $[\hat\vartheta(x),\hat\rho(y)]=\ii\delta(x-y)$, Hamilton's equations generated by Eq.~\eqref{eq:HIR} read
\begin{subequations}\label{eq:eom}
\begin{align}
  \partial_t\hat\vartheta
  &=\chi^{-1}\hat\rho+ag_\rho\hat m,
  \label{eq:eom_theta}\\
  \partial_t\hat\rho+\partial_x\hat j
  &=-g_\theta\hat m,
  \qquad
  \hat j=-\rho_s\partial_x\hat\vartheta .
  \label{eq:eom_rho}
\end{align}
\end{subequations}
Note that the second equation agrees with the leading expansion of the exact lattice
source $\hat{\mathcal S}_i=\ii\lambda_z\nm_i(\hat a_i-\ad_i)$. Eliminating
either field gives
\begin{subequations}\label{eq:waves}
\begin{align}
  (\partial_t^2-v^2\partial_x^2)\hat\vartheta
  &=ag_\rho\,\partial_t\hat m-\chi^{-1}g_\theta\hat m,
  \label{eq:wave_theta}\\
  (\partial_t^2-v^2\partial_x^2)\hat\rho
  &=a\rho_sg_\rho\,\partial_x^2\hat m
  -g_\theta\,\partial_t\hat m,
  \label{eq:wave_rho}
\end{align}
\end{subequations}
where $v^2=\rho_s\chi^{-1}$. The retarded Green's function
\begin{equation}\label{eq:GR}
  G_R(x,t)=\frac{1}{2v}\Theta(t)\Theta(vt-|x|)
\end{equation}
restricts the response to the past sound cone. The leading edges of a detached long-wavelength disturbance therefore propagate at $v$, independently of the source velocity. For $\theta_0=0$, a stationary source in an adiabatically dressed fluid has the quasistatic deformation
\begin{equation}\label{eq:rho_qs}
  \rho_{\mathrm{qs}}=-\chi ag_\rho m .
\end{equation}
For a slowly moving source, this approximates the near field that follows the
pair; velocity-dependent corrections grow as the pair approaches the sound
velocity. In our protocol, the fluid is initially undeformed and the coupling is
switched on at $t=0$. The mismatch with Eq.~\eqref{eq:rho_qs} excites a
transient even for a stationary profile. For a moving rigid packet,
$m(x,t)=m_0(x-X_0-ut)$ and
$\partial_tm=-u\,\partial_xm_0$, producing an additional drive. The response
can thus be organized as
\begin{equation}\label{eq:response_decomposition}
  \delta\rho
  \simeq
  \rho_{\mathrm{qs}}
  +\delta\rho_{\mathrm{switch}}
  +\delta\rho_{\mathrm{motion}} .
\end{equation}
Figure~\ref{fig:diagram} summarizes the attached contribution and the retarded
fronts propagating at $\pm v$, whose amplitudes depend on the preparation and
source trajectory.

For direct comparison with the many-body data, we retain the lattice theory and use the measured magnon density as an external source. With $a=1$ and
$\tau=U_Bt$,
\begin{subequations}\label{eq:lattice_eom}
\begin{align}
  \partial_\tau\vartheta_i
  &=\Pi_i+\tilde g_\rho m_i(\tau),
  \label{eq:lattice_theta}\\
  \partial_\tau\Pi_i
  &=2r\nbar(\vartheta_{i+1}-2\vartheta_i+\vartheta_{i-1})
  -\tilde g_\theta m_i(\tau),
  \label{eq:lattice_Pi}
\end{align}
\end{subequations}
with
\begin{equation}\label{eq:gtilde}
  \tilde g_\theta=-2\eta\sqrt{\nbar}\sin\theta_0,
  \qquad
  \tilde g_\rho=\frac{\eta}{\sqrt{\nbar}}\cos\theta_0 .
\end{equation}
Here, $m_i(\tau)=\avg{\nm_i(\tau)}$ is measured in the full simulation. It incorporates the coupling-dependent position, width, and velocity of the pair at the one-body level, while neglecting magnon-density fluctuations and connected spin-boson correlations. The field $\Pi_i$ is compared with $\delta n_i(\tau)=\avg{\hat n_i(\tau)}-\avg{\hat n_i(0)}$. Boundary conditions and initial data are specified in Appendix~\ref{app:response}. For $\theta_0=0$, one has $\tilde g_\theta=0$, so that summing Eq.~\eqref{eq:lattice_Pi} over the chain and imposing no current through the boundaries gives
\begin{equation}\label{eq:linear_number_balance}
  \partial_\tau\delta N_{\mathrm{hyd}}
  =0,
  \qquad
  \delta N_{\mathrm{hyd}}=\sum_i\Pi_i .
\end{equation}
Thus, the magnon pair redistributes the density through the potential
$\tilde g_\rho m_i$, but the linear hydrodynamic theory conserves the total
boson number to the retained order. The lattice current is
$j_{i+1/2}=-2r\nbar(\vartheta_{i+1}-\vartheta_i)$, and removing the external
source gives
\begin{equation}\label{eq:lattice_disp}
  \tilde\omega_q
  =2\sqrt{2r\nbar}\left|\sin\frac q2\right|
  \xrightarrow[q\to0]{}
  \sqrt{2r\nbar}\,|q| .
\end{equation}
Short-wavelength components travel below the linear sound-cone edge and
contribute to the thickness and trailing structure of the fronts. For the
intermediate Gross--Pitaevskii description, we set
$\psi_i=\sqrt{\nbar+\Pi_i}\,e^{-\ii\vartheta_i}$ and evolve
\begin{equation}\label{eq:discrete_gp}
\begin{aligned}
  \ii\partial_\tau\psi_i
  &=-r_{\mathrm{eff}}(\psi_{i-1}+\psi_{i+1}) \\
  &\quad+\big(|\psi_i|^2-s_{\mathrm{eff}}\big)\psi_i
  +\eta m_i(\tau),
\end{aligned}
\end{equation}
where
$r_{\mathrm{eff}}=c_{\mathrm{TDVP}}^2/(2\nbar)$,
$s_{\mathrm{eff}}=\nbar-2r_{\mathrm{eff}}$, and $c_{\mathrm{TDVP}}$ is the
dimensionless front velocity measured at the comparison point. The cone slope
is therefore matched by construction, while its amplitude and structure remain
predictions. Unlike the linear equation, Eq.~\eqref{eq:discrete_gp} does not
conserve $N_{\mathrm{GP}}=\sum_i|\psi_i|^2$:
\begin{equation}\label{eq:gp_number_balance}
\begin{aligned}
  \partial_\tau N_{\mathrm{GP}}
  &=-2\eta\sum_i m_i(\tau)\Imm\psi_i
  \\
  &=2\eta\sum_i m_i(\tau)
  \sqrt{\nbar+\Pi_i}\sin\vartheta_i .
\end{aligned}
\end{equation}
The additive displacement source can therefore create and destroy bosons. For an
initially real field, the rate vanishes at $\tau=0$, but the induced phase makes
the number response begin at $O(\eta^2)$, consistent with the exact
microscopic source derived in Appendix~\ref{app:sb}. Averaging
$|\psi_i^{(\phi)}(\tau)|^2-|\psi_i^{(\phi)}(0)|^2$ over uniform global phases
represents the $U(1)$-symmetric initial condition semiclassically, although it
does not reproduce all fixed-number fluctuations. Linearization recovers
Eq.~\eqref{eq:lattice_eom} with $\theta_0=0$ and
$r\to r_{\mathrm{eff}}$.

The construction just presented treats the measured average magnon density as a classical
source and linearizes the hydrodynamic fields. It therefore captures the
sound-cone geometry, front velocity, and weak-response scaling, while
neglecting source fluctuations, connected spin-boson correlations, and
nonlinear amplitudes. To go beyond this treatment and test these approximations, we compare these results with the full
many-body evolution in the following section.

%% file: sec_v3/04_numerics.tex
\section{Numerical results}\label{sec:num}

In this section, we compare the effective descriptions of Section~\ref{sec:hydro} with the full
lattice dynamics in three steps. We first characterize how the Bose--Hubbard
medium modifies the motion and spatial profile of the two-magnon bound
state. We then identify the parameter regime in which the bosonic response
develops detached fronts and compare their propagation velocity with the
semiclassical sound velocity. Finally, we compare the linear hydrodynamic and
nonlinear semiclassical approximations with the full TDVP evolution and
verify the exact boson-number balance equation. Unless stated otherwise, the
simulations use open chains of $L=80$ composite sites, $s=0.5$, $K_0=\pi/2$,
and evolve up to $\tau=U_Bt=30$, before boundary reflections reach the
central region.

\subsection{Dynamics of the two-magnon bound state}\label{sec:num_pair}

\begin{figure}[h!]
  \centering
  \includegraphics[width=0.9\columnwidth]{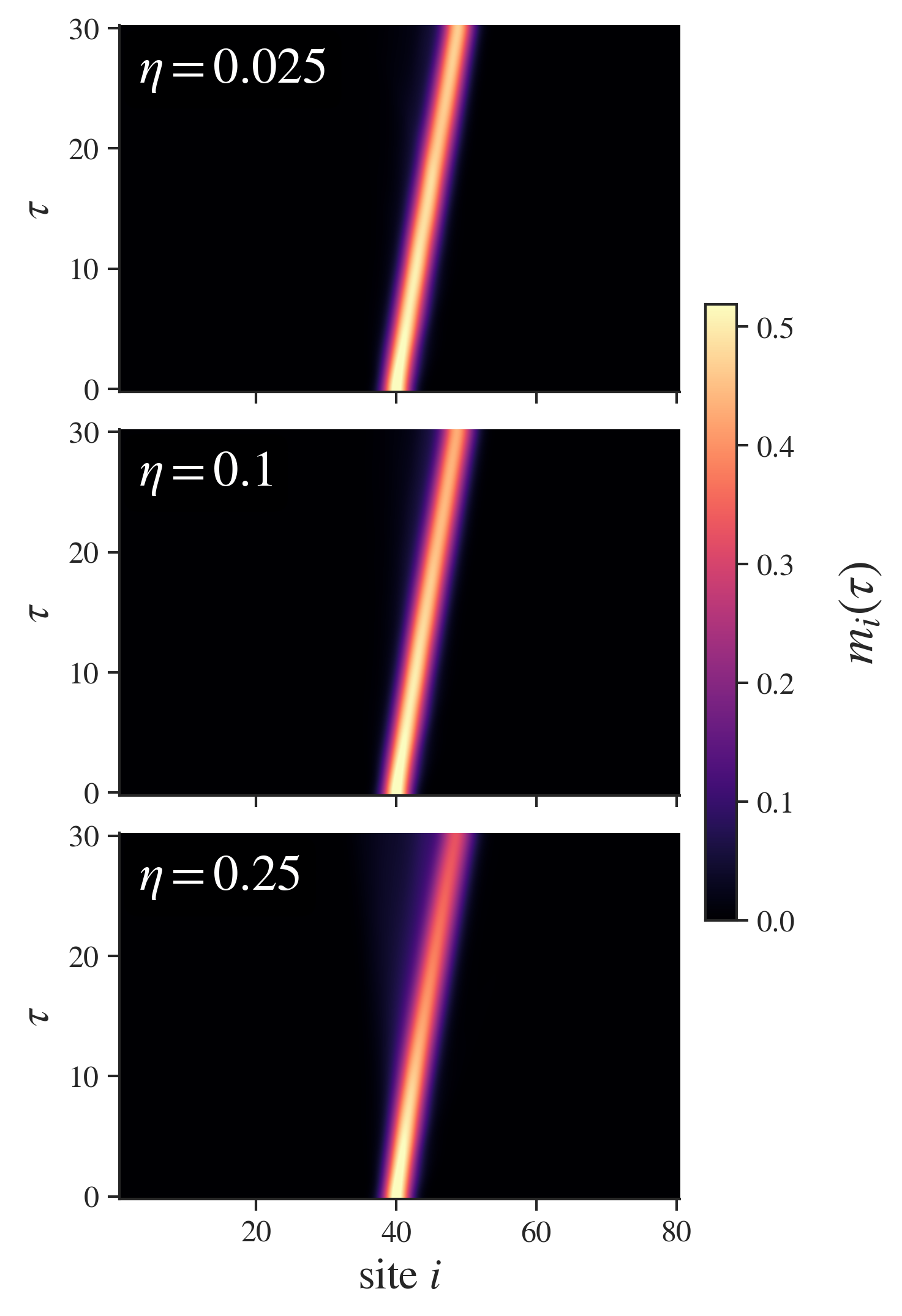}
\caption{Real-time magnon density $m_i(\tau)$ for $\eta=0.025$, $0.1$, and $0.25$ at $K_0=\pi/2$. The Bose--Hubbard parameters
are fixed at $r=0.40$ and $s=0.5$, within the compressible phase. Increasing the spin-boson coupling slows and broadens the bound pair
and produces a weak trailing tail, while the excitation remains spatially
localized over the simulated interval.}
  \label{fig:source_rt}
\end{figure}

The hydrodynamic construction uses the magnon density as the source that drives
the fluid, whereas in the full Hamiltonian the pair is itself dynamical and is
modified by the medium. Figure~\ref{fig:source_rt} shows the real-time magnon density
for three values of $\eta$, with the bosonic parameters fixed in the
compressible phase. At weak coupling, the pair follows a narrow,
approximately straight trajectory close to that of the isolated XXZ chain.
Increasing $\eta$ reduces the slope of this trajectory and broadens the
profile, which also develops a weak trailing tail. The pair nevertheless
remains a single localized excitation throughout the displayed time window. 

\begin{figure}[h!]
  \centering
  \includegraphics[width=0.5\textwidth]{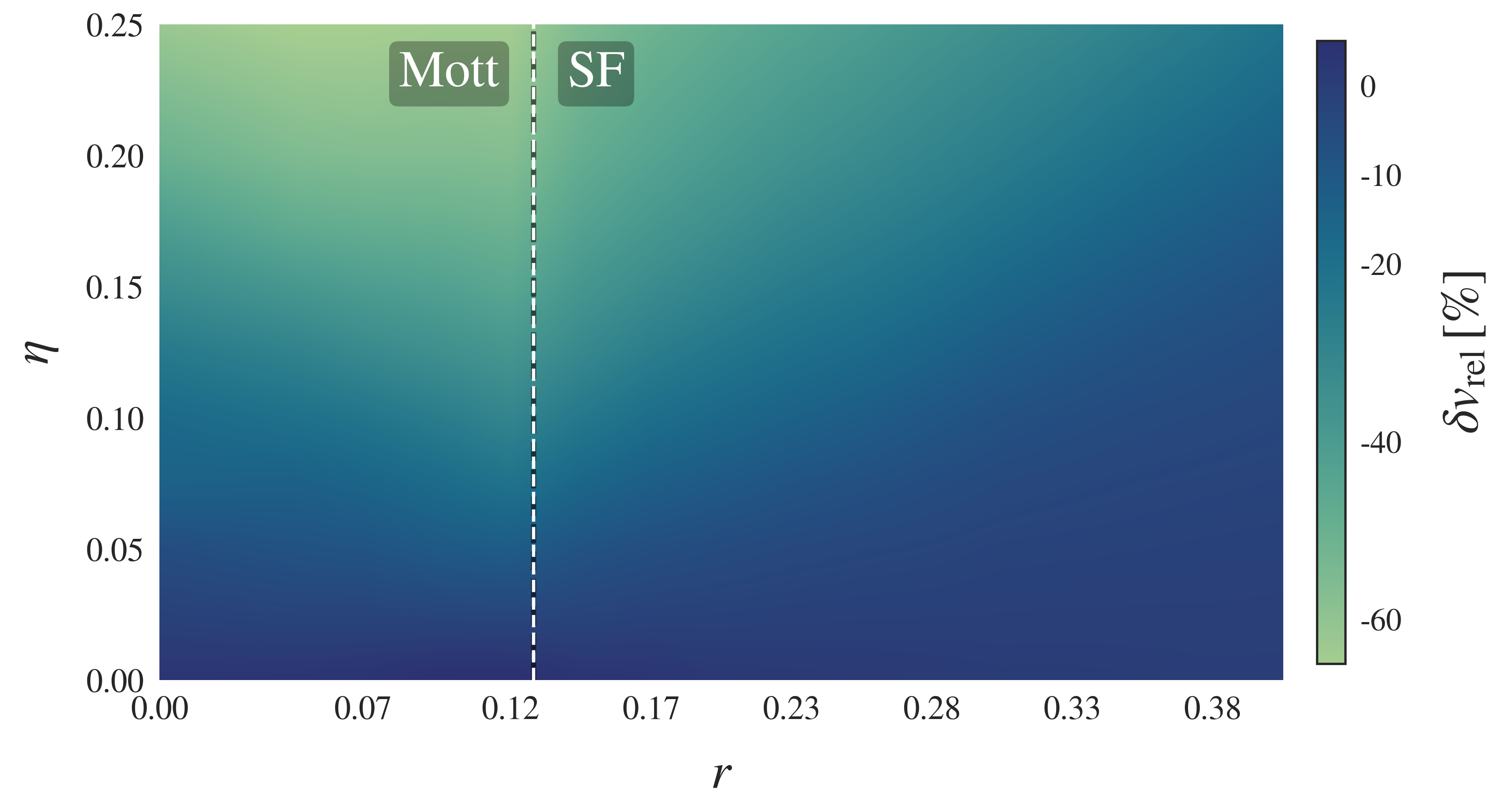}
  \caption{Relative velocity shift of the two-magnon bound state over the
  $(r,\eta)$ plane, defined in Eq.~\eqref{eq:dv_rel}. Negative values denote a
  reduction with respect to the matched uncoupled XXZ packet. The vertical
  dashed line marks the equilibrium Mott--superfluid boundary along $s=0.5$.}
  \label{fig:backreaction_velocity}
\end{figure}

To quantify the change in the translational motion of the pair, we define the center of
mass
\begin{equation}\label{eq:magnon_com}
  X_{\mathrm m}(\tau)=\frac12\sum_i i\,\avg{\nm_i(\tau)}
\end{equation}
and extract $v_{\mathrm{SB}}$ from a linear fit over the propagation window. A
matched calculation with $\eta=0$, using the same trapped state, kick, and fit
interval, gives the reference velocity $v_{\mathrm{XXZ}}$. Their relative
difference is
\begin{equation}\label{eq:dv_rel}
  \delta v_{\mathrm{rel}}
  =\frac{v_{\mathrm{SB}}-v_{\mathrm{XXZ}}}
  {|v_{\mathrm{XXZ}}|}\;[\%].
\end{equation}
As shown in Fig.~\ref{fig:backreaction_velocity},
$\delta v_{\mathrm{rel}}$ is negative over nearly the entire parameter plane.
The velocity reduction is small at weak coupling but grows rapidly with
$\eta$, reaching its largest values at the highest couplings. Its smooth
variation across the equilibrium Mott--superfluid boundary indicates that this
slowing is a dressing effect on the mobile pair rather than a diagnostic of the
bosonic phase transition. 

\begin{figure}[h!]
  \centering
  \includegraphics[width=0.5\textwidth]{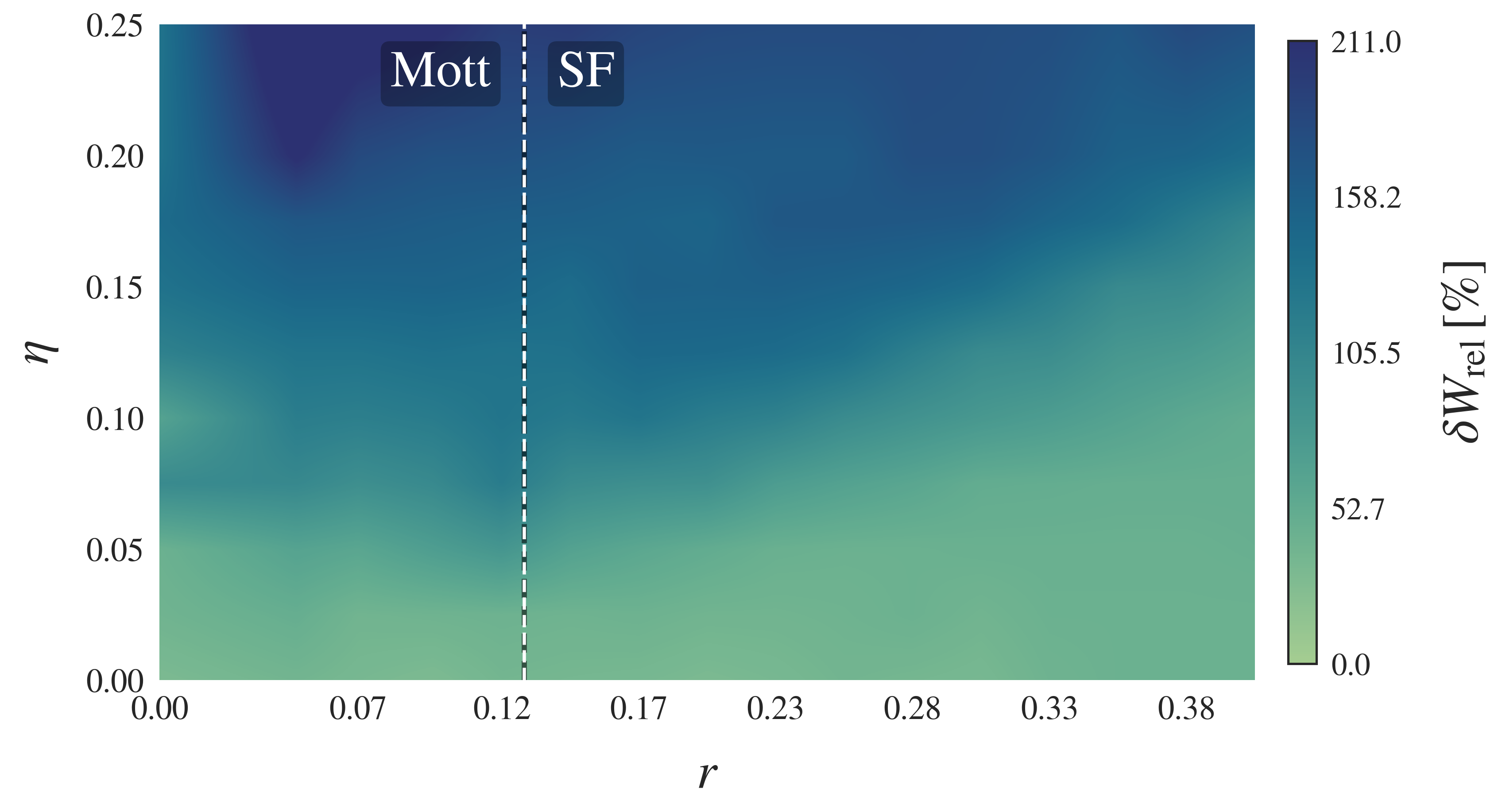}
  \caption{Relative change in the thresholded spatial width of the main
  magnon-density packet over the $(r,\eta)$ plane, as defined in
  Eq.~\eqref{eq:dW_rel}. The packet broadens progressively as the coupling increases,
  while the dependence remains smooth across the Mott--superfluid boundary
  marked by the vertical dashed line.}
  \label{fig:backreaction_size}
\end{figure}

We quantify the spatial broadening of the pair directly from the one-body magnon-density profile. At each time, we retain the sites satisfying $m_i(\tau)\geq0.05\max_j m_j(\tau)$ and select the connected component that contains the density maximum. Denoting its outermost sites by
$x_{\mathrm L}(\tau)$ and $x_{\mathrm R}(\tau)$, the thresholded packet width is $W_{\mathrm p}(\tau)=x_{\mathrm R}(\tau)-x_{\mathrm L}(\tau)$. The early and late values are averages over the first and last $20\%$, respectively, of the interval $1\leq\tau\leq29$. Their relative change is
\begin{equation}\label{eq:dW_rel}
  \delta W_{\mathrm{rel}}
  =\frac{W_{\mathrm p}^{\mathrm{late}}
  -W_{\mathrm p}^{\mathrm{early}}}
  {W_{\mathrm p}^{\mathrm{early}}}\;[\%].
\end{equation}
Figure~\ref{fig:backreaction_size} shows that the spatial broadening follows the
same coupling dependence as the slowing, remaining modest at weak coupling but
becoming pronounced at large $\eta$, where the thresholded packet width can more than triple. The joint evolution of $X_{\mathrm m}$ and $W_{\mathrm p}$ characterizes the source that acts on the fluid. In the comparisons below, we therefore use the measured density $m_i(\tau)$, which automatically includes the coupling-dependent position, velocity, and spatial profile of the pair. The absence of a sharp feature at the Mott--superfluid boundary also makes clear that the
transition is revealed primarily by the propagation of the bosonic response,
not by a singular change in the bound state itself.

\subsection{Emergence of the hydrodynamic sound cone}
\label{sec:num_hydro}

Having characterized the backreaction on the bound pair, we now turn to the
response of the bosonic medium. Along $s=0.5$, the isolated Bose--Hubbard chain
leaves the unit-filling Mott lobe near $r\simeq0.13$
(Appendix~\ref{app:bh}, Fig.~\ref{fig:phase}). The equilibrium ground states
also provide the background density $\nbar(r,s)$ that enters the semiclassical
prediction in Eq.~\eqref{eq:vsound}. We characterize the induced signal by
\begin{equation}\label{eq:density_response}
  \delta n_i(\tau)
  =\avg{\hat n_i(\tau)}-\avg{\hat n_i(0)},
\end{equation}
which removes the weak inhomogeneity of the initial open-chain profile.

\begin{figure}[h!]
  \centering
  \includegraphics[width=\columnwidth]{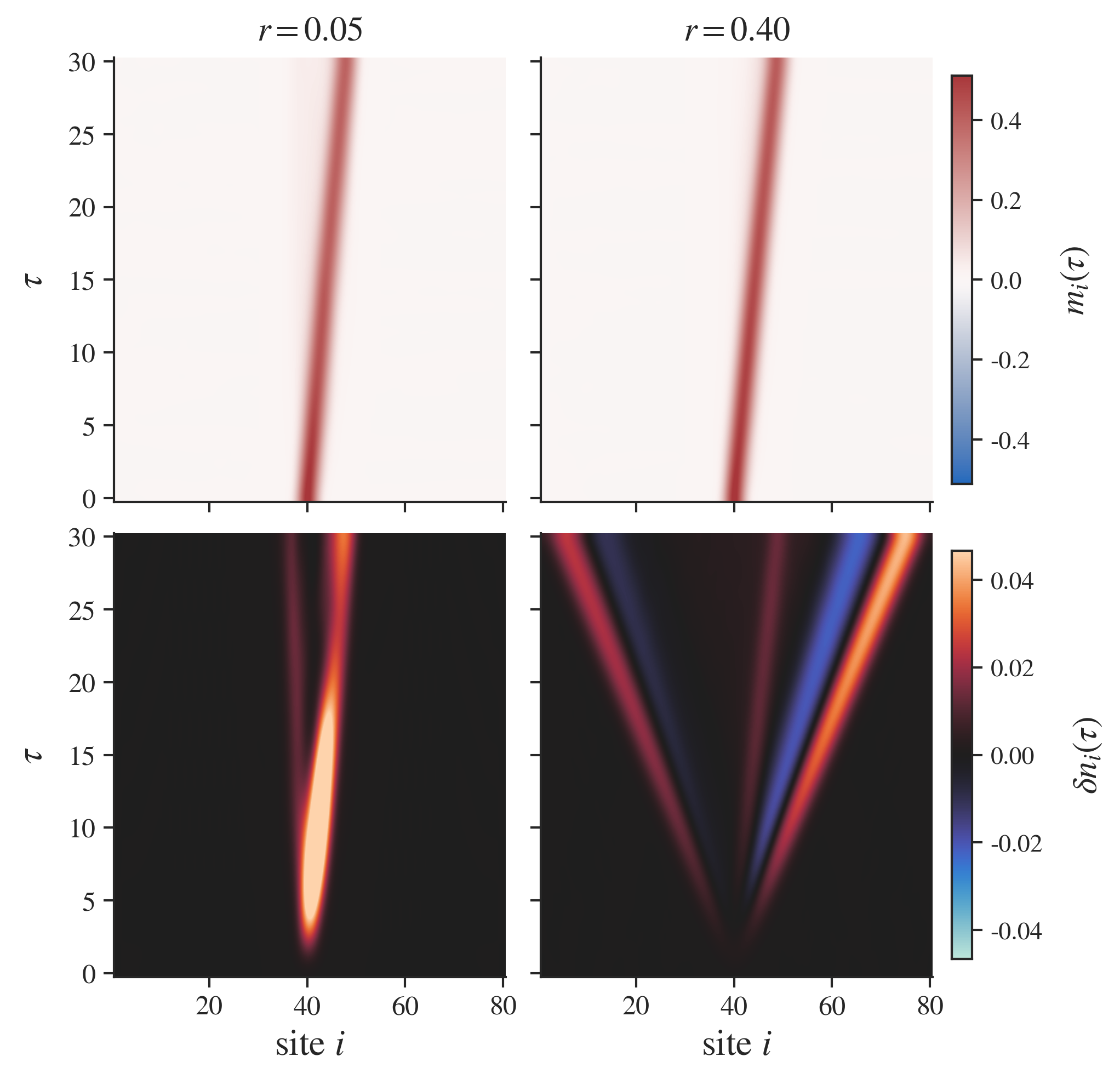}
\caption{Space-time dynamics across the Mott--superfluid transition at fixed
spin-boson coupling $\eta=0.1$. Columns correspond to $r=0.05$ and $r=0.40$.
The upper row shows the magnon density $m_i(\tau)$; the lower row shows the
induced bosonic density $\delta n_i(\tau)$. In the Mott regime, the response
remains localized around the pair, whereas in the compressible regime two
counterpropagating fronts detach and form a sound cone.}
  \label{fig:evolution}
\end{figure}

Figure~\ref{fig:evolution} compares a point inside the Mott lobe,
$r=0.05$, with a point deep in the compressible phase, $r=0.40$, at the same
spin-boson coupling. In the Mott regime, the pair moves across the chain while
the induced bosonic deformation remains concentrated around it. In the
compressible regime, by contrast, two fronts detach from the release region
and propagate in opposite directions, while a narrower component continues to
follow the pair. Each detached signal displays a compression-rarefaction
structure, consistent with the derivative source in
Eq.~\eqref{eq:wave_rho}. The fronts form the sound cone predicted by the
retarded response in Eq.~\eqref{eq:GR}, and their finite width reflects both the spatial
extent of the source and the bending of the lattice dispersion in
Eq.~\eqref{eq:lattice_disp} away from its long-wavelength linear form.

\begin{figure}[t]
  \centering
  \includegraphics[width=\columnwidth]{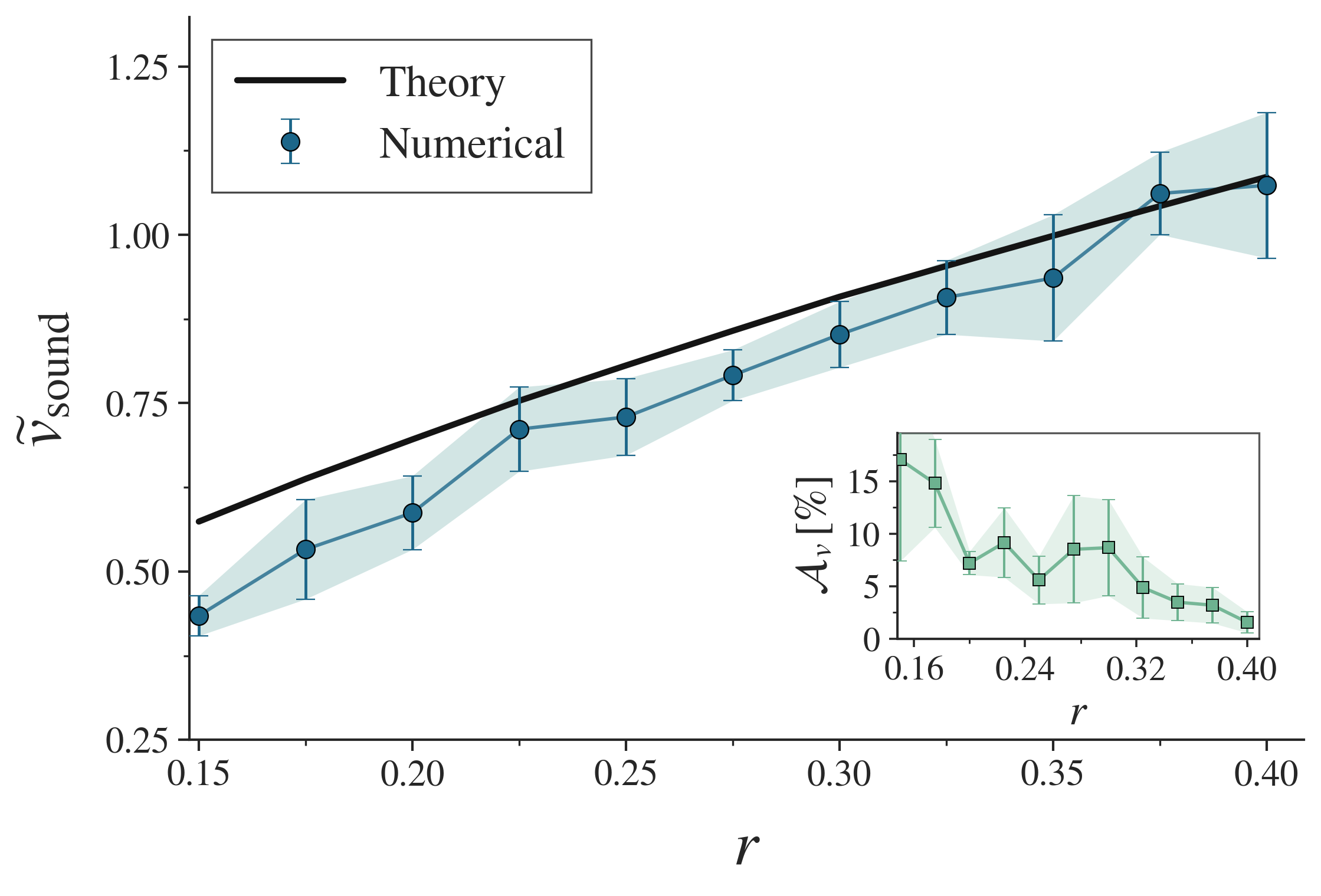}
\caption{Velocity of the detached density fronts in the compressible phase.
Markers show the numerical sound-front velocity defined in
Eq.~\eqref{eq:front_velocity}, using $a=U_B=1$; bars and the shaded region
represent the directional and ridge-fit spread. The solid line is the
semiclassical prediction from Eq.~\eqref{eq:vsound}, evaluated using the equilibrium background density and without fitted parameters. The inset shows the directional velocity asymmetry normalized by the
semiclassical sound velocity, as defined in Eq.~\eqref{eq:front_asymmetry}.}
  \label{fig:sound}
\end{figure}

We determine the velocity of these fronts by masking the source region, tracking
the left- and right-moving ridges before boundary reflections arrive, and
defining
\begin{subequations}\label{eq:front_observables}
\begin{align}
  \tilde v_{\mathrm{sound}}^{\mathrm{num}}
  &=\frac{|\tilde v_+|+|\tilde v_-|}{2},
  \qquad
  \tilde v_\pm=\frac{v_\pm}{aU_B},
  \label{eq:front_velocity}\\
\mathcal A_v
&=\frac{\big||\tilde v_+|-|\tilde v_-|\big|}
{2\,\vsc/(aU_B)}\;[\%].
\label{eq:front_asymmetry}
\end{align}
\end{subequations}
Figure~\ref{fig:sound} compares the measured mean velocity with
$\vsc/(aU_B)=\sqrt{2r\nbar}$, evaluated using the equilibrium density and no
fitted parameters. The numerical front velocity increases with $r$ and approaches
the semiclassical prediction deeper in the compressible phase. The larger
difference close to the Mott boundary is consistent with the stronger quantum
renormalization of the compressibility and phase stiffness in that region.
The theory-normalized left-right velocity asymmetry shown in the inset decreases as the fronts separate more clearly from the moving source. The common velocity scale of the two fronts, despite
their opposite propagation directions, identifies them as signatures of a collective mode of the fluid rather than of the pair velocity.

\subsection{Comparison of dynamical descriptions}
\label{sec:num_comparison}

\begin{figure*}[t]
  \centering
  \includegraphics[width=\textwidth]{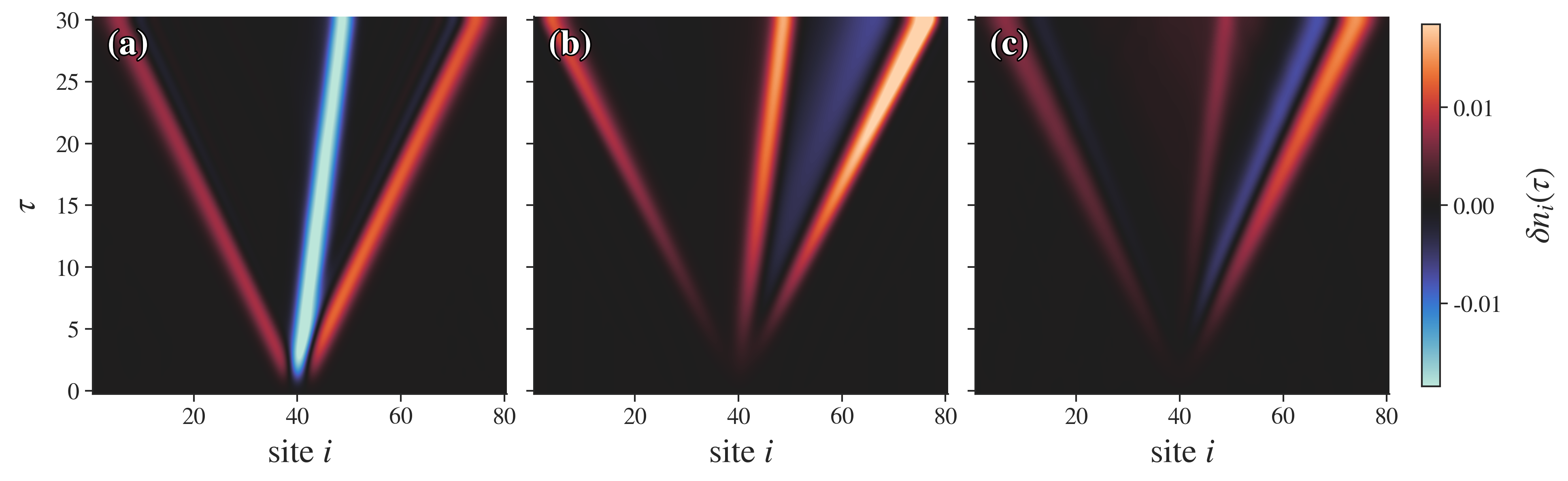}
  \caption{Bosonic density response for $L=80$, $r=0.40$, $s=0.5$, and
  $\eta=0.05$. (a) Linear hydrodynamics from Eq.~\eqref{eq:lattice_eom}.
  (b) Nonlinear Gross--Pitaevskii dynamics, averaged over initial global
  phases. (c) TDVP evolution of the full Hamiltonian in Eq.~\eqref{eq:H}.
  Panels~(a) and (b) use the same TDVP magnon density as an external source and
  the same velocity-matched $r_{\mathrm{eff}}$; all panels share a common
  color scale.}
  \label{fig:hydrodynamic_comparison}
\end{figure*}

To assess how the observed response is captured at different levels of approximation, we compare the three descriptions in Fig.~\ref{fig:hydrodynamic_comparison} at $r=0.40$, $s=0.5$, and $\eta=0.05$. Panels~(a) and (b) are driven by the same measured TDVP profile,
$m_i(\tau)=\avg{\hat S_i^z+1/2}_{\mathrm{TDVP}}$, so differences among the panels isolate the treatment of the bosonic sector rather than changes in the trajectory of the source. In both semiclassical calculations, $r_{\mathrm{eff}}=c_{\mathrm{TDVP}}^2/(2\nbar)$ fixes the sound-cone slope to the measured TDVP value; the amplitude and detailed spatial structure remain predictions.

Panel~(a) integrates the linear lattice equations in Eq.~\eqref{eq:lattice_eom} for
$\theta_0=0$. It reproduces the detached fronts and their propagation
velocity, but at this order the magnon density acts only as a local potential.
Consequently,
$\partial_\tau\sum_i\Pi_i=0$ as stated in
Eq.~\eqref{eq:linear_number_balance}, and the linear theory cannot reproduce a
net change in the total boson number. Panel~(b) instead integrates the nonlinear Gross--Pitaevskii equation in
Eq.~\eqref{eq:discrete_gp} and averages over initial global phases. As described
by the number balance in Eq.~\eqref{eq:gp_number_balance}, its additive source
allows the total boson number to change from $O(\eta^2)$ onward. This nonlinear
description reproduces the relative amplitudes and attached structure more
accurately than the linear approximation. Panel~(c) is the TDVP evolution of
the full Hamiltonian in Eq.~\eqref{eq:H}, in which boson-number change, quantum
fluctuations, and spin-boson entanglement are all retained. The remaining
differences between panels~(b) and (c) reflect these many-body effects together
with the limitations of the semiclassical parametrization.

\begin{figure}[t]
  \centering
  \includegraphics[width=\columnwidth]{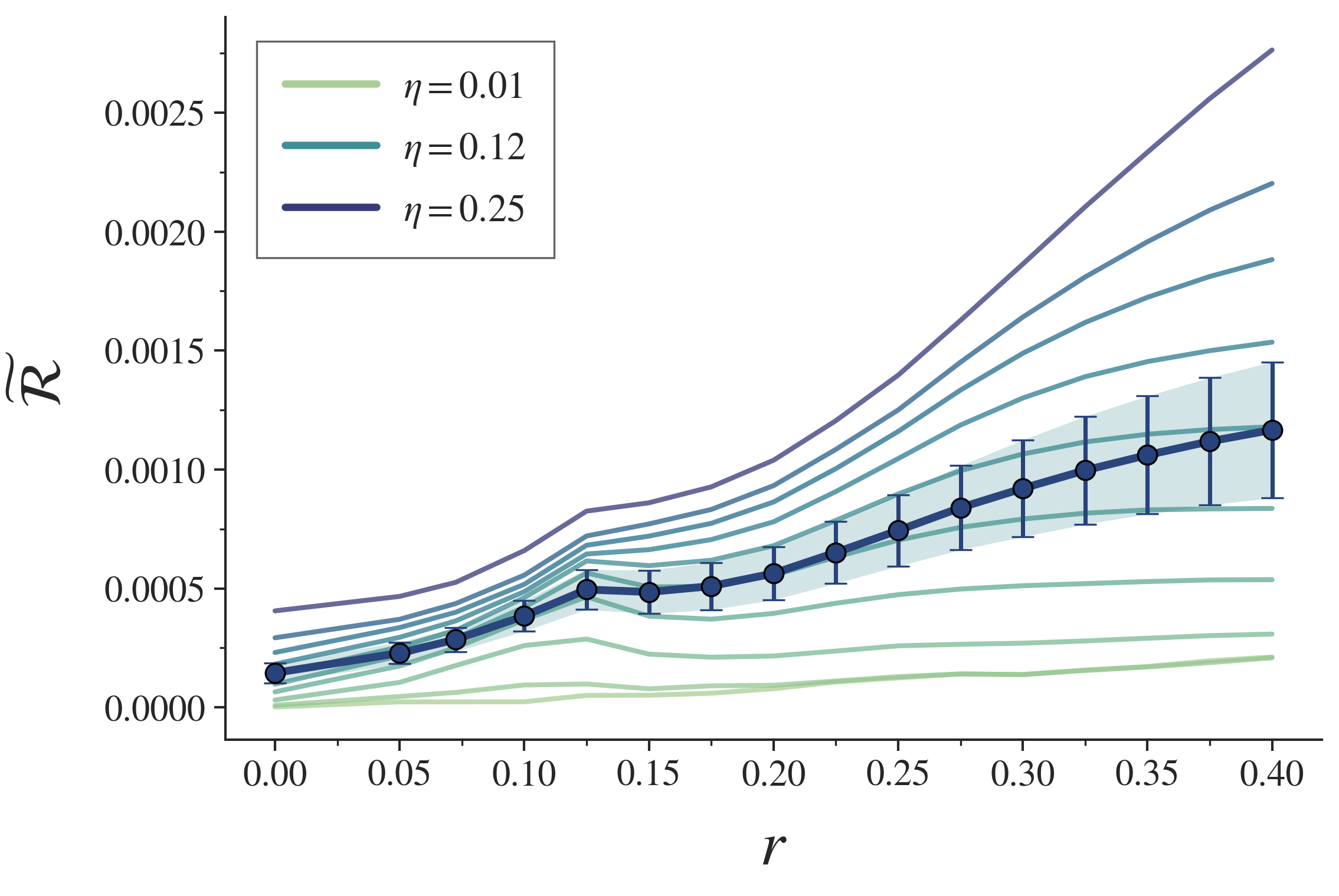}
  \caption{Dimensionless root-mean-square residual
  $\widetilde{\mathcal R}$ of the exact local balance equation in
  Eq.~\eqref{eq:balance}. Thin curves show the scan in spin-boson coupling, with
  representative values indicated in the legend; markers and the shaded band
  summarize the central value and spread across the scan. The residual stays small throughout
  the parameter range, providing an independent consistency check of the TDVP
  trajectories and measured observables.}
  \label{fig:continuity}
\end{figure}

Because the displacement interaction creates and destroys bosons, the relevant
microscopic identity is the continuity equation with its local source,
\begin{equation}\label{eq:balance}
  \partial_t\avg{\hat n_i}
  +\avg{\hat j_{i+1/2}}-\avg{\hat j_{i-1/2}}
  =\avg{\hat{\mathcal S}_i},
\end{equation}
where
\begin{subequations}\label{eq:current_source}
\begin{align}
  \hat j_{i+1/2}
  &=-\ii t_B
  \big(\ad_i\hat a_{i+1}-\ad_{i+1}\hat a_i\big),
  \label{eq:current}\\
  \hat{\mathcal S}_i
  &=\ii\lambda_z\nm_i\big(\hat a_i-\ad_i\big).
  \label{eq:source}
\end{align}
\end{subequations}
This relation follows directly from the full Hamiltonian and must hold
independently of the hydrodynamic reduction. Using the stored TDVP trajectories,
we evaluate the local residual
\begin{subequations}\label{eq:balance_residual}
\begin{align}
  \mathcal R_i(t)
  &=\partial_t n_i+j_{i+1/2}-j_{i-1/2}-\mathcal S_i,
  \label{eq:local_residual}\\
  \mathcal R(t)
  &=
  \left[
  \frac{1}{L}
  \sum_{i=1}^{L}|\mathcal R_i(t)|^2
  \right]^{1/2},
  \label{eq:rms_residual}
\end{align}
\end{subequations}
using backward finite differences between consecutive stored frames. The
finite-difference derivative, and hence $\mathcal R(t)$, is defined only from
the second stored frame onward. Here
$n_i=\avg{\hat n_i}$, $j_{i+1/2}=\avg{\hat j_{i+1/2}}$, and
$\mathcal S_i=\avg{\hat{\mathcal S}_i}$. Because the chain is open, no current
flows through its ends, $j_{1/2}=j_{L+1/2}=0$, so that
Eq.~\eqref{eq:local_residual} is defined at every site and the sum in
Eq.~\eqref{eq:rms_residual} runs over the whole chain.
Figure~\ref{fig:continuity} shows the
temporal average of $\mathcal R(t)$ over the available stored frames. The dimensionless residual
$\widetilde{\mathcal R}=\mathcal R/U_B$ remains in the range
$10^{-4}$--$10^{-3}$ across most of the scan and increases smoothly as
hopping and coupling make the evolution numerically more demanding. The small residual in
Fig.~\ref{fig:continuity} shows that the density, current, and microscopic
source extracted from TDVP satisfy the same local balance to high accuracy.
The source $\hat{\mathcal S}_i$ accounts for the local boson-number change
generated by the displacement coupling. This process is present in the full
TDVP and nonlinear Gross--Pitaevskii dynamics, but is absent to leading order
from the linear hydrodynamic theory at $\theta_0=0$.

%% file: sec_v3/05_discussion.tex
\section{Discussion and conclusions}\label{sec:disc}

We derived a long-wavelength description of the response of a Bose--Hubbard
fluid to a mobile two-magnon bound state and compared it with the full lattice
dynamics. The reduced Hamiltonian contains a density channel, $g_\rho$, which
acts as a local chemical-potential shift, and a phase channel, $g_\theta$,
which changes the local boson number. For the real background considered here,
$g_\theta=0$ at linear order, so the leading theory conserves the total boson
number, while number-changing processes reappear in the nonlinear and
microscopic dynamics. The simulations show that the fluid progressively slows
and broadens the bound pair without destroying its localized character. In the
Mott phase, the bosonic deformation remains attached to the moving source,
whereas in the compressible phase two counterpropagating fronts detach and
travel at speeds approaching the semiclassical sound velocity in
Eq.~\eqref{eq:vsound}. The linear theory reproduces the geometry and
propagation velocity of these fronts, while the nonlinear equation more
accurately captures their amplitudes, the attached component, and the net
boson-number response. The small residual of the exact local balance equation
provides an independent consistency check of the TDVP dynamics.

These results separate the evolution of the source from the collective
response of the fluid. Increasing the coupling modifies the velocity and
thresholded spatial width of the magnon-density packet, as shown in
Figs.~\ref{fig:backreaction_velocity} and \ref{fig:backreaction_size}, but the sound
cone remains governed by the properties of the compressible medium. Using the
measured magnon density as the source incorporates the evolving position,
velocity, and width of the pair, although it neglects magnon-density
fluctuations and connected spin-boson correlations. Quantitative differences
remain because the parameters in Eq.~\eqref{eq:chirhos} are semiclassical
estimates and are most strongly renormalized near the Mott transition. In
addition, the linear reduction does not describe large-amplitude fluctuations,
while the nonlinear phase-averaged field does not reproduce all correlations
of the fixed-number quantum state. These limitations affect amplitudes and
local structure more strongly than the positions and velocities of the
detached fronts.

The ingredients of the model are accessible across several quantum-simulation
platforms. Optical lattices realize Bose--Hubbard fluids tunable across the
Mott--superfluid transition and permit the observation of density
dynamics~\cite{Jaksch1998,Greiner2002,Cheneau2012}; quantum-gas microscopes
resolve mobile magnons and their bound states~\cite{Fukuhara2013,
FukuharaMobile2013}; superconducting circuit-QED arrays can realize interacting
bosonic modes through microwave resonators with transmon-induced Kerr
nonlinearities~\cite{Hartmann2016,SchmidtKoch2013}; and Rydberg-dressed arrays
and trapped ions offer spin-boson couplings~\cite{Plodzien2018,Kosior2023,
Mallick2025}. The present setting combines these ingredients so that the
composite quasiparticle and the fluid act as mutual probes: the pair excites
the collective response, while its velocity and spatial profile record the
influence of the medium. Finally, our setting is closely related to
nuclear physics studies of high-energy impurity propagation through strongly
coupled colored matter; see,
e.g.,~\cite{Ruppert:2005uz,Casalderrey-Solana:2006lmc,
Casalderrey-Solana:2004fdk,Satarov:2005mv,Yang:2022nei,
Casalderrey-Solana:2016jvj,BarataRico2026}. Establishing a direct connection
requires extending the model to higher dimensions with local gauge symmetries,
which we leave for future work.

%% file: sec_v3/06_acknowledgments.tex
\begin{acknowledgments}
M.P., A.N.C., and A.R.\ acknowledge computational resources provided by the Spanish Supercomputing
Network (RES) at the Barcelona
Supercomputing Center on MareNostrum 5 under allocation NNO-2025-3-0004. This work has been partially funded by the Eric \& Wendy Schmidt Fund for Strategic Innovation through the CERN Next Generation Triggers project under grant agreement number SIF-2023-004.\\
\end{acknowledgments}

\textbf{Data Availability:} Data are available from the authors upon
reasonable request.

%% file: sec_v3/appendix_bose_hubbard.tex
\section{Density-phase expansion of the Bose--Hubbard chain}
\label{app:bh}

This appendix derives the reduction summarized in Section~\ref{sec:hydro},
from the microscopic Hamiltonian in Eq.~\eqref{eq:HBH} to the continuum form
in Eq.~\eqref{eq:Hhyd}, and records the semiclassical parameters together with
their regime of validity.

\subsection{Quadratic density-phase expansion}

We write the occupation as a background plus a fluctuation and introduce the
conjugate phase,
\begin{equation}\label{eq:appA_rotor}
  \hat n_i=\nbar+\hat\Pi_i,
  \qquad
  \hat a_i\simeq e^{-\ii\hat\theta_i}\sqrt{\nbar+\hat\Pi_i},
  \qquad
  \ad_i\simeq\sqrt{\nbar+\hat\Pi_i}\,e^{\ii\hat\theta_i},
\end{equation}
with $[\hat\theta_i,\hat\Pi_j]=\ii\delta_{ij}$. The representation in
Eq.~\eqref{eq:appA_rotor} is not an operator identity; it is the standard
long-wavelength rotor description, accurate when the coarse-grained density
fluctuation is small relative to the background density and the phase varies
slowly from site to site. Using $\hat n_i(\hat n_i-1)=\hat n_i^2-\hat n_i$, the local part of
Eq.~\eqref{eq:HBH} is
\begin{equation}\label{eq:appA_loc1}
  \frac{U_B}{2}\hat n_i(\hat n_i-1)-\mu\hat n_i
  =\frac{U_B}{2}\hat n_i^2
  -\left(\frac{U_B}{2}+\mu\right)\hat n_i .
\end{equation}
Substituting $\hat n_i=\nbar+\hat\Pi_i$ and expanding,
\begin{align}
  \frac{U_B}{2}\hat n_i^2
  &=\frac{U_B}{2}\nbar^2
  +U_B\nbar\hat\Pi_i
  +\frac{U_B}{2}\hat\Pi_i^2,
  \label{eq:appA_quad}\\
  -\left(\frac{U_B}{2}+\mu\right)\hat n_i
  &=-\left(\frac{U_B}{2}+\mu\right)\nbar
  -\left(\frac{U_B}{2}+\mu\right)\hat\Pi_i .
  \label{eq:appA_lin}
\end{align}
Combining these contributions and discarding the $\hat\Pi$-independent block,
\begin{equation}\label{eq:appA_loc2}
  \hat H_{\mathrm{loc}}
  =\mathrm{const.}
  +\sum_i
  \left[
    \left(U_B\nbar-\frac{U_B}{2}-\mu\right)\hat\Pi_i
    +\frac{U_B}{2}\hat\Pi_i^2
  \right].
\end{equation}
The local term is therefore exactly quadratic in the density fluctuation, with a linear coefficient that vanishes at $\mu=U_B(\nbar-\tfrac12)$. This is the atomic-limit relation between chemical potential and filling; it is modified by hopping, as shown below. Using Eq.~\eqref{eq:appA_rotor}, defining
$\delta\hat\theta_i=\hat\theta_{i+1}-\hat\theta_i$, and neglecting local
amplitude-phase commutators at the semiclassical order of the rotor
description, we obtain the bond operator
\begin{equation}\label{eq:appA_hop_rotor}
  \ad_i\hat a_{i+1}+\hc
  =2\sqrt{\nbar+\hat\Pi_i}\sqrt{\nbar+\hat\Pi_{i+1}}
  \cos\big(\delta\hat\theta_i\big).
\end{equation}
Expanding the square roots with
$\sqrt{1+x}\simeq1+\tfrac x2-\tfrac{x^2}8$,
\begin{equation}\label{eq:appA_sqrt}
  \sqrt{\nbar+\hat\Pi_i}
  \simeq\sqrt{\nbar}
  \left[
    1+\frac{\hat\Pi_i}{2\nbar}-\frac{\hat\Pi_i^2}{8\nbar^2}
  \right],
\end{equation}
and multiplying the two factors gives
\begin{equation}\label{eq:appA_sqrtprod}
  \sqrt{\nbar+\hat\Pi_i}\sqrt{\nbar+\hat\Pi_{i+1}}
  \simeq
  \nbar
  +\frac{\hat\Pi_i+\hat\Pi_{i+1}}{2}
  -\frac{\big(\hat\Pi_{i+1}-\hat\Pi_i\big)^2}{8\nbar},
\end{equation}
where we used
$\tfrac{\hat\Pi_i\hat\Pi_{i+1}}{4\nbar}
-\tfrac{\hat\Pi_i^2+\hat\Pi_{i+1}^2}{8\nbar}
=-\tfrac{(\hat\Pi_i-\hat\Pi_{i+1})^2}{8\nbar}$.
Expanding the cosine as $1-\tfrac12(\delta\hat\theta_i)^2$ and dropping the
product of $\hat\Pi$ with $(\delta\hat\theta)^2$, which is cubic in the
fluctuations,
\begin{equation}\label{eq:appA_bond}
  \ad_i\hat a_{i+1}+\hc
  \simeq
  2\nbar
  +\big(\hat\Pi_i+\hat\Pi_{i+1}\big)
  -\nbar\big(\delta\hat\theta_i\big)^2
  -\frac{\big(\hat\Pi_{i+1}-\hat\Pi_i\big)^2}{4\nbar}.
\end{equation}
Multiplying by $-t_B$ and summing over bonds,
\begin{align}
  \hat H_{\mathrm{hop}}
  &\simeq\mathrm{const.}
  -t_B\sum_i\big(\hat\Pi_i+\hat\Pi_{i+1}\big)
  \nonumber\\
  &\quad
  +t_B\nbar\sum_i\big(\hat\theta_{i+1}-\hat\theta_i\big)^2
  +\frac{t_B}{4\nbar}\sum_i\big(\hat\Pi_{i+1}-\hat\Pi_i\big)^2 .
  \label{eq:appA_hop}
\end{align}
The four terms are, in order, a constant; a linear density term that in the
bulk of a long chain equals $-2t_B\sum_i\hat\Pi_i$; the phase-stiffness term;
and a density-gradient term. The last term becomes
$\propto(\partial_x\hat\rho)^2$ in
the continuum and carries two derivatives more than $\hat\rho^2$, so it is
irrelevant in the infrared and is dropped from the leading hydrodynamic theory.
It does contribute curvature to the dispersion at finite momentum, alongside the
lattice curvature retained in Eq.~\eqref{eq:lattice_disp}.

\subsection{Hydrodynamic reduction and effective parameters}

Adding the linear pieces of Eqs.~\eqref{eq:appA_loc2} and \eqref{eq:appA_hop}
gives the total coefficient of $\sum_i\hat\Pi_i$, in units of $U_B$,
\begin{equation}\label{eq:appA_Atot}
  \frac{A_{\mathrm{tot}}}{U_B}
  =\nbar-\frac12-s-2r .
\end{equation}
Expanding about a stationary background requires $A_{\mathrm{tot}}=0$, giving
\begin{equation}\label{eq:appA_eos}
  s=\nbar-\frac12-2r .
\end{equation}
Equation~\eqref{eq:appA_eos} is the rotor-level equation of state. In the
one-dimensional quantum model, the background density is instead the
thermodynamic function $\nbar(r,s)$ of the interacting ground state, which is
what we use in Section~\ref{sec:num}; the role of Eq.~\eqref{eq:appA_eos} in the derivation is only to guarantee that the expansion is taken about a stationary point, so that no linear term survives to shift the fields. Dropping constants, imposing $A_{\mathrm{tot}}=0$, and discarding the
density-gradient term leaves the lattice infrared Hamiltonian
\begin{equation}\label{eq:appA_lat}
  \hat H_{\mathrm{B}}^{\mathrm{IR}}
  =\frac{U_B}{2}\sum_i\hat\Pi_i^2
  +t_B\nbar\sum_i\big(\hat\theta_{i+1}-\hat\theta_i\big)^2,
\end{equation}
which, when divided by $U_B$, gives
$\tfrac12\sum_i\Pi_i^2+r\nbar\sum_i(\theta_{i+1}-\theta_i)^2$, whose equations
of motion are given in Eq.~\eqref{eq:lattice_eom}. Passing to
the continuum with $x_i=ia$,
\begin{equation}\label{eq:appA_cont_rules}
  \hat\Pi_i=a\hat\rho(x_i),
  \qquad
  \hat\theta_{i+1}-\hat\theta_i\simeq a\partial_x\hat\theta,
  \qquad
  \sum_i\simeq\frac1a\int\dd x,
\end{equation}
the two terms become
\begin{align}
  \frac{U_B}{2}\sum_i\hat\Pi_i^2
  &\simeq\frac{U_Ba}{2}\int\dd x\,\hat\rho^2,
  \label{eq:appA_cont1}\\
  t_B\nbar\sum_i\big(\hat\theta_{i+1}-\hat\theta_i\big)^2
  &\simeq t_B\nbar a\int\dd x\,\big(\partial_x\hat\theta\big)^2,
  \label{eq:appA_cont2}
\end{align}
so that comparing with Eq.~\eqref{eq:Hhyd} gives
\begin{equation}\label{eq:appA_params}
  \chi^{-1}=U_Ba,
  \qquad
  \rho_s=2t_B\nbar a .
\end{equation}
Retaining the gradient term would add
$\tfrac{t_Ba^3}{4\nbar}\int\dd x\,(\partial_x\hat\rho)^2$ to
Eq.~\eqref{eq:Hhyd}.

The Hamiltonian in Eq.~\eqref{eq:Hhyd} yields a wave equation with
$v^2=\rho_s\chi^{-1}$, hence
\begin{equation}\label{eq:appA_v}
  \vsc=a\sqrt{2U_Bt_B\nbar}
  =U_Ba\sqrt{2r\nbar},
  \qquad
  \tilde v\equiv\frac{\vsc}{aU_B}=\sqrt{2r\nbar}.
\end{equation}
To connect with the standard Luttinger form
\begin{equation}\label{eq:appA_LL}
  \hat H_{\mathrm{LL}}
  =\frac{v}{2\pi}\int\dd x
  \left[
    \mathcal K\big(\partial_x\hat\theta\big)^2
    +\frac{1}{\mathcal K}\big(\partial_x\hat\varphi\big)^2
  \right],
  \qquad
  \hat\rho=\frac1\pi\partial_x\hat\varphi ,
\end{equation}
we substitute $\partial_x\hat\varphi=\pi\hat\rho$, which turns
Eq.~\eqref{eq:appA_LL} into
$\tfrac12\int\dd x[\tfrac{\pi v}{\mathcal K}\hat\rho^2
+\tfrac{v\mathcal K}{\pi}(\partial_x\hat\theta)^2]$. A term-by-term comparison
with Eq.~\eqref{eq:Hhyd} gives
\begin{equation}\label{eq:appA_LLrel}
  \chi^{-1}=\frac{\pi v}{\mathcal K},
  \qquad
  \rho_s=\frac{v\mathcal K}{\pi},
  \qquad
  v^2=\rho_s\chi^{-1},
  \qquad
  \mathcal K=\pi\sqrt{\rho_s\chi},
\end{equation}
and substituting Eq.~\eqref{eq:appA_params} then gives
\begin{equation}\label{eq:appA_K}
  \mathcal K_{\mathrm{sc}}=\pi\sqrt{2r\nbar}.
\end{equation}
Equations~\eqref{eq:appA_v} and \eqref{eq:appA_K} are semiclassical. In one
dimension, particularly at $\nbar\simeq1$ near the Mott lobe, both $v$ and
$\mathcal K$ are renormalized by quantum
fluctuations~\cite{Haldane1981,Kuhner1998,Ejima2012}; the functional form
in Eq.~\eqref{eq:Hhyd} survives throughout the superfluid phase but its
coefficients must then be read as effective parameters.

\subsection{Regime of validity and equilibrium reference}

The hydrodynamic reduction assumes a stationary, compressible, phase-stiff
background, long-wavelength disturbances satisfying $qa\ll1$, and fluctuations
sufficiently small for the quadratic truncation of
Eqs.~\eqref{eq:appA_loc2}--\eqref{eq:appA_bond} to hold. Its applicability
therefore changes qualitatively across the Mott--superfluid transition, which
reflects the competition between hopping and on-site repulsion. In the Mott
phase at integer filling, the interaction suppresses fluctuations in the local
boson occupation, the compressibility
$\chi=a^{-1}\partial\nbar/\partial\mu$ vanishes, and the low-energy spectrum
consists of gapped particle-hole excitations. A perturbation cannot launch a
low-energy acoustic wave in this regime, and the induced density response remains predominantly localized or is carried by gapped excitations.

\begin{figure}[h!]
  \centering
  \includegraphics[width=0.5\textwidth]{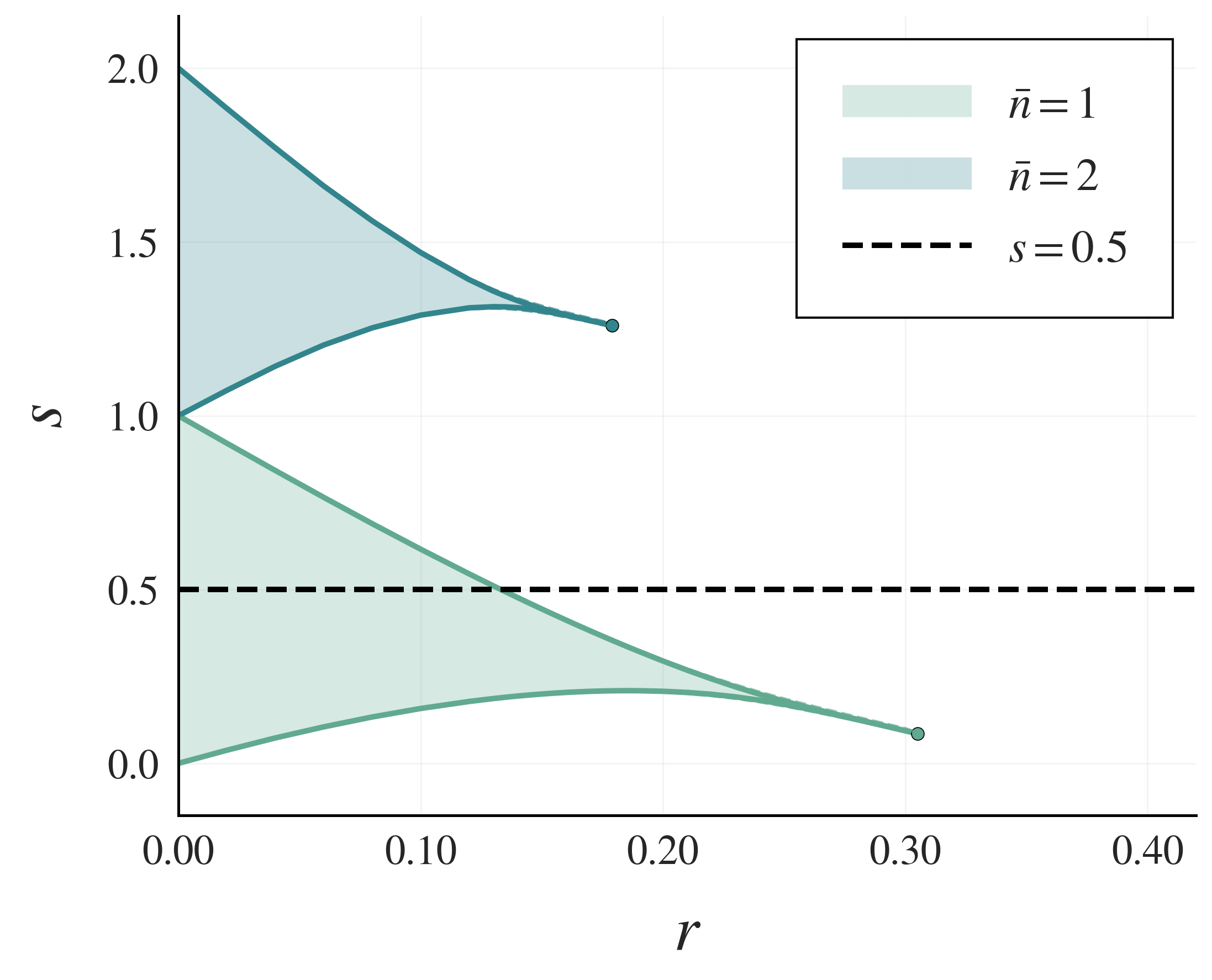}
\caption{Equilibrium reference for the dynamical scan. Shaded regions denote the $\nbar=1$ and $\nbar=2$ Mott lobes of the one-dimensional Bose--Hubbard model in the $(r,s)$ plane. The dashed line shows the scan trajectory $s=0.5$, which leaves the unit-filling lobe near $r\simeq0.13$. The Mott region is incompressible and has gapped particle-hole excitations, whereas the superfluid region is compressible and supports a gapless acoustic mode.}
  \label{fig:phase}
\end{figure}

On the superfluid side, hopping delocalizes the bosons and enhances fluctuations
in the local boson occupation, producing a compressible, phase-stiff background
with a gapless density-phase mode. The semiclassical theory is therefore
expected to reproduce the long-wavelength sector of the many-body dynamics,
including the propagation of detached density fronts. Quantitative deviations
can nevertheless arise near the transition, where quantum fluctuations
strongly renormalize the effective parameters, and for short-wavelength or
large-amplitude disturbances. Figure~\ref{fig:phase} locates the scan used in
Section~\ref{sec:num} relative to this boundary. The line $s=0.5$ leaves the
unit-filling lobe near $r\simeq0.13$, so agreement with the hydrodynamic
description is expected only on the compressible side of this crossing. The
growth of $\nbar$ along the same line at fixed $s$ makes the semiclassical
velocity in Eq.~\eqref{eq:appA_v} increase faster than $\sqrt r$.

\vspace{7cm}

%% file: sec_v3/appendix_spin_boson.tex
\section{Hydrodynamic expansion of the spin-boson coupling}
\label{app:sb}

This appendix derives the source couplings, the quadratic phase term that survives for a real background, and the exact lattice source operator whose leading expansion fixes the sign of $g_\theta$. The microscopic interaction is
\begin{equation}\label{eq:appB_HSB}
  \HSB=\lambda_z\sum_i\nm_i\big(\hat a_i+\ad_i\big),
  \qquad
  \nm_i=\Sz_i+\half ,
\end{equation}
so that the bosonic field is driven only where the magnon density is nonzero. Using Eq.~\eqref{eq:appA_rotor} with the symmetrized semiclassical ordering specified in Appendix~\ref{app:bh}, we obtain
\begin{equation}\label{eq:appB_disp}
  \hat a_i+\ad_i
  \simeq
  \sqrt{\nbar+\hat\Pi_i}
  \left(e^{-\ii\hat\theta_i}+e^{\ii\hat\theta_i}\right)
  =2\sqrt{\nbar+\hat\Pi_i}\,\cos\hat\theta_i ,
\end{equation}
which is accurate to the same order as the rotor representation
itself. Expanding about the background phase,
$\hat\theta_i=\theta_0+\hat\vartheta_i$, using
\begin{equation}\label{eq:appB_expansions}
  \sqrt{\nbar+\hat\Pi_i}
  \simeq\sqrt{\nbar}\left(1+\frac{\hat\Pi_i}{2\nbar}\right),
  \qquad
  \cos\big(\theta_0+\hat\vartheta_i\big)
  \simeq\cos\theta_0-\sin\theta_0\,\hat\vartheta_i ,
\end{equation}
and retaining only terms linear in the hydrodynamic fields, thereby discarding
the product $\hat\Pi_i\hat\vartheta_i$, gives
\begin{equation}\label{eq:appB_linear}
  \HSB\simeq
  2\lambda_z\sqrt{\nbar}\cos\theta_0\sum_i\nm_i
  -2\lambda_z\sqrt{\nbar}\sin\theta_0\sum_i\nm_i\hat\vartheta_i
  +\frac{\lambda_z}{\sqrt{\nbar}}\cos\theta_0\sum_i\nm_i\hat\Pi_i .
\end{equation}
The first term contains no bosonic fluctuation. In a sector of fixed magnon number it is a constant, and if the magnon number were not fixed it would act as a chemical potential for magnons, that is, a renormalization of the longitudinal field $h_z$ in Eq.~\eqref{eq:HXXZ}. The remaining two terms define
\begin{equation}\label{eq:appB_gs}
  g_\theta=-2\lambda_z\sqrt{\nbar}\sin\theta_0,
  \qquad
  g_\rho=\frac{\lambda_z}{\sqrt{\nbar}}\cos\theta_0,
\end{equation}
with dimensionless counterparts
$\tilde g_\theta=g_\theta/U_B=-2\eta\sqrt{\nbar}\sin\theta_0$ and
$\tilde g_\rho=g_\rho/U_B=(\eta/\sqrt{\nbar})\cos\theta_0$. In terms of the
continuum magnon density in Eq.~\eqref{eq:mcont}, the lattice sums become
\begin{equation}\label{eq:appB_cont}
  \sum_i\nm_i\hat\vartheta_i=\int\dd x\,\hat m\hat\vartheta,
  \qquad
  \sum_i\nm_i\hat\Pi_i=a\int\dd x\,\hat m\hat\rho ,
\end{equation}
and these relations turn Eq.~\eqref{eq:appB_linear} into Eq.~\eqref{eq:Hsrc}. Note the opposite scaling of the two channels with the background density: the phase source grows as $\sqrt{\nbar}$, whereas the density potential is suppressed as $1/\sqrt{\nbar}$, because a given displacement corresponds to a smaller relative density change in a denser fluid. We next specialize the expansion to the real background used throughout the numerical calculations. For $\theta_0=0$, one has $\sin\theta_0=0$ and
$\cos\theta_0=1$, hence $g_\theta=0$ and
$g_\rho=\lambda_z/\sqrt{\nbar}$, and
Eq.~\eqref{eq:appB_linear} reduces to
\begin{equation}\label{eq:appB_real}
  \hat H_{\mathrm{SB}}^{\mathrm{lin}}
  =\frac{\lambda_z}{\sqrt{\nbar}}\sum_i\nm_i\hat\Pi_i
  =a\frac{\lambda_z}{\sqrt{\nbar}}\int\dd x\,\hat m\hat\rho .
\end{equation}
The vanishing of the linear phase source does not make the microscopic coupling number-conserving; Eq.~\eqref{eq:appB_HSB} manifestly changes the boson number by one. This means only that the number source vanishes at first order in the response about a real background. Carrying the cosine expansion one order further, $\cos\hat\vartheta_i\simeq1-\tfrac12\hat\vartheta_i^2$, gives the first surviving phase term
\begin{equation}\label{eq:appB_quad}
  \hat H_{\mathrm{SB}}^{(2,\theta)}
  =-\lambda_z\sqrt{\nbar}\sum_i\nm_i\hat\vartheta_i^2 ,
\end{equation}
which becomes relevant once the phase fluctuations around the magnon are no
longer small. Since Eq.~\eqref{eq:appB_quad} is quadratic in
$\hat\vartheta$ and localized around the magnon, it acts as a local softening
of the phase sector rather than as a drive and becomes increasingly relevant
in the strong-coupling region in which the magnon-density packet broadens, as shown in
Fig.~\ref{fig:backreaction_size}. More precisely, the induced phase obeys $\hat\vartheta=O(\lambda_z)$ at leading order, so the contribution of Eq.~\eqref{eq:appB_quad} to the linearized equation of motion is $O(\lambda_z^2)$. Dropping it is therefore justified here by the expansion to first order in $\lambda_z$, not merely by counting powers of the hydrodynamic fields. An independent consistency check follows from the exact local continuity equation. Without invoking the hydrodynamic expansion, the spin-boson contribution is
\begin{equation}\label{eq:appB_exact}
  \partial_t\hat n_i\big|_{\mathrm{SB}}
  =\ii\big[\HSB,\hat n_i\big]
  =\ii\lambda_z\nm_i\big(\hat a_i-\ad_i\big)
  \equiv\hat{\mathcal S}_i ,
\end{equation}
which is Eq.~\eqref{eq:source} and is the quantity measured in
Fig.~\ref{fig:continuity}. Applying the same semiclassical rotor ordering,
\begin{equation}\label{eq:appB_sinth}
  \ii\big(\hat a_i-\ad_i\big)
  \simeq\ii\sqrt{\nbar+\hat\Pi_i}
  \left(e^{-\ii\hat\theta_i}-e^{\ii\hat\theta_i}\right)
  =2\sqrt{\nbar+\hat\Pi_i}\,\sin\hat\theta_i ,
\end{equation}
so that
\begin{equation}\label{eq:appB_source_rotor}
  \hat{\mathcal S}_i
  \simeq2\lambda_z\nm_i\sqrt{\nbar+\hat\Pi_i}\,\sin\hat\theta_i .
\end{equation}
Expanding $\sin(\theta_0+\hat\vartheta_i)\simeq\sin\theta_0
+\cos\theta_0\,\hat\vartheta_i$ gives
\begin{equation}\label{eq:appB_source_leading}
  \hat{\mathcal S}_i\simeq
  2\lambda_z\sqrt{\nbar}
  \big(\sin\theta_0+\cos\theta_0\,\hat\vartheta_i\big)\nm_i .
\end{equation}
The field-independent contribution equals $-g_\theta\nm_i$, reproduces the
right-hand side of Eq.~\eqref{eq:eom_rho}, and fixes the sign of $g_\theta$ in
Eq.~\eqref{eq:appB_gs}. For $\theta_0=0$ the remaining term is
$2\lambda_z\sqrt{\nbar}\,\nm_i\hat\vartheta_i$; because
$\hat\vartheta_i=O(\lambda_z)$, it affects the induced response at
$O(\lambda_z^2)$ and is consistently absent from the leading-order theory.
This agreement provides a consistency check on the whole construction, since
the same microscopic term that supplies the number-changing source in the
hydrodynamic continuity equation must also be retained to close the exact
lattice balance law in Eq.~\eqref{eq:balance}.

%% file: sec_v3/appendix_response.tex
\section{Hydrodynamic response and lattice equations}
\label{app:response}

This appendix derives the driven equations quoted in
Section~\ref{sec:hydro}, gives their retarded solution, and specifies the lattice form used for direct comparison with the many-body dynamics.

\subsection{Continuum equations of motion}

Combining Eqs.~\eqref{eq:Hhyd} and \eqref{eq:Hsrc} gives
\begin{equation}\label{eq:appC_H}
  \hat H_{\mathrm{IR}}
  =\frac12\int\dd x
  \left[
    \chi^{-1}\hat\rho^2+\rho_s\big(\partial_x\hat\vartheta\big)^2
  \right]
  +g_\theta\!\int\!\dd x\,\hat m\hat\vartheta
  +ag_\rho\!\int\!\dd x\,\hat m\hat\rho ,
\end{equation}
and, using $[\hat\vartheta(x),\hat\rho(y)]=\ii\delta(x-y)$, Hamilton's equations
take the form
\begin{equation}\label{eq:appC_var}
  \partial_t\hat\vartheta=\frac{\delta\hat H_{\mathrm{IR}}}{\delta\hat\rho},
  \qquad
  \partial_t\hat\rho=-\frac{\delta\hat H_{\mathrm{IR}}}{\delta\hat\vartheta}.
\end{equation}
The density derivative is immediate,
$\delta\hat H_{\mathrm{IR}}/\delta\hat\rho=\chi^{-1}\hat\rho+ag_\rho\hat m$.
For the phase, integration by parts, with the boundary terms discarded, gives
\begin{equation}\label{eq:appC_byparts}
\begin{aligned}
  \delta\!\left[\frac12\int\dd x\,\rho_s
  \big(\partial_x\hat\vartheta\big)^2\right]
  &=\rho_s\!\int\!\dd x\,\big(\partial_x\hat\vartheta\big)
  \big(\partial_x\delta\hat\vartheta\big)\\
  &=-\rho_s\!\int\!\dd x\,\big(\partial_x^2\hat\vartheta\big)
  \delta\hat\vartheta ,
\end{aligned}
\end{equation}
so that
$\delta\hat H_{\mathrm{IR}}/\delta\hat\vartheta
=-\rho_s\partial_x^2\hat\vartheta+g_\theta\hat m$. Therefore,
\begin{subequations}\label{eq:appC_eom}
\begin{align}
  \partial_t\hat\vartheta
  &=\chi^{-1}\hat\rho+ag_\rho\hat m,
  \label{eq:appC_theta}\\
  \partial_t\hat\rho
  &=\rho_s\partial_x^2\hat\vartheta-g_\theta\hat m .
  \label{eq:appC_rho}
\end{align}
\end{subequations}
Defining the current $\hat j=-\rho_s\partial_x\hat\vartheta$, so that
$\partial_x\hat j=-\rho_s\partial_x^2\hat\vartheta$,
Eq.~\eqref{eq:appC_rho} becomes the sourced continuity equation
\begin{equation}\label{eq:appC_cont}
  \partial_t\hat\rho+\partial_x\hat j=-g_\theta\hat m ,
\end{equation}
which is Eq.~\eqref{eq:eom_rho}. The same expression follows from the
microscopic current: with
$\ad_i\hat a_{i+1}\simeq\nbar e^{-\ii(\theta_{i+1}-\theta_i)}$ one has $\Imm\avg{\ad_i\hat a_{i+1}}\simeq-\nbar(\theta_{i+1}-\theta_i)$, hence $j_{i+1/2}=2t_B\Imm\avg{\ad_i\hat a_{i+1}}
\simeq-2t_B\nbar a\,\partial_x\theta=-\rho_s\partial_x\vartheta$, using Eq.~\eqref{eq:appA_params} and $\partial_x\theta=\partial_x\vartheta$. The coupled first-order equations can be recast as driven wave equations. Solving Eq.~\eqref{eq:appC_theta} for the density gives
\begin{equation}\label{eq:appC_rhosol}
  \hat\rho=\chi\big(\partial_t\hat\vartheta-ag_\rho\hat m\big),
\end{equation}
and differentiating once in time and substituting into
Eq.~\eqref{eq:appC_cont} then yields
\begin{equation}\label{eq:appC_sub}
  \chi\big(\partial_t^2\hat\vartheta-ag_\rho\partial_t\hat m\big)
  -\rho_s\partial_x^2\hat\vartheta=-g_\theta\hat m .
\end{equation}
Dividing by $\chi$ and using $v^2=\rho_s\chi^{-1}$ gives
\begin{equation}\label{eq:appC_wave_theta}
  \big(\partial_t^2-v^2\partial_x^2\big)\hat\vartheta
  =ag_\rho\,\partial_t\hat m-\chi^{-1}g_\theta\hat m .
\end{equation}
Eliminating $\hat\vartheta$ instead by differentiating
Eq.~\eqref{eq:appC_rho} in time and using Eq.~\eqref{eq:appC_theta} gives the equivalent density equation, and the current obeys a first-order relation:
\begin{subequations}\label{eq:appC_others}
\begin{align}
  \big(\partial_t^2-v^2\partial_x^2\big)\hat\rho
  &=a\rho_sg_\rho\,\partial_x^2\hat m-g_\theta\,\partial_t\hat m,
  \label{eq:appC_wave_rho}\\
  \partial_t\hat j+v^2\partial_x\hat\rho
  &=-a\rho_sg_\rho\,\partial_x\hat m .
  \label{eq:appC_current}
\end{align}
\end{subequations}
The two source channels involve different derivatives and are therefore distinguishable in principle: the number-changing source enters through $\partial_t m$, whereas the density potential enters through $\partial_x^2m$. In terms of the dimensionless variables used in the main text, with $\tilde v=\vsc/(aU_B)=\sqrt{2r\nbar}$,
Eq.~\eqref{eq:appC_wave_theta} reads
\begin{equation}\label{eq:appC_wave_dimless}
  \big(\partial_\tau^2-a^2\tilde v^2\partial_x^2\big)\hat\vartheta
  =a\frac{\eta}{\sqrt{\nbar}}\cos\theta_0\,\partial_\tau\hat m
  +2a\eta\sqrt{\nbar}\sin\theta_0\,\hat m ,
\end{equation}
and for the real background $\theta_0=0$ the wave equations reduce to
\begin{equation}\label{eq:appC_wave_real}
  \big(\partial_\tau^2-a^2\tilde v^2\partial_x^2\big)\hat\vartheta
  =a\frac{\eta}{\sqrt{\nbar}}\,\partial_\tau\hat m,
  \qquad
  \big(\partial_\tau^2-a^2\tilde v^2\partial_x^2\big)\hat\rho
  =2a^2r\nbar\frac{\eta}{\sqrt{\nbar}}\,\partial_x^2\hat m .
\end{equation}

\subsection{Retarded solution and the sound cone}

Writing Eq.~\eqref{eq:appC_wave_theta} as
$(\partial_t^2-v^2\partial_x^2)\vartheta=J_{\mathrm{eff}}$ with
\begin{equation}\label{eq:appC_Jeff}
  J_{\mathrm{eff}}(x,t)
  =ag_\rho\,\partial_tm(x,t)-\chi^{-1}g_\theta\,m(x,t),
\end{equation}
the retarded Green's function is defined by
\begin{equation}\label{eq:appC_Gdef}
  \big(\partial_t^2-v^2\partial_x^2\big)G_R(x,t)=\delta(x)\delta(t),
  \qquad
  G_R(x,t<0)=0,
\end{equation}
and in one spatial dimension equals
\begin{equation}\label{eq:appC_G}
  G_R(x,t)=\frac{1}{2v}\Theta(t)\,\Theta\big(vt-|x|\big).
\end{equation}
The response is therefore
\begin{equation}\label{eq:appC_sol}
  \vartheta(x,t)
  =\vartheta_{\mathrm{hom}}(x,t)
  +\int_{-\infty}^{t}\!\!\dd t'\!\int\!\dd x'\,
  G_R(x-x',t-t')\,J_{\mathrm{eff}}(x',t'),
\end{equation}
whose support is restricted by the step functions to the past sound cone
$|x-x'|\le v(t-t')$. Two consequences are used in the main text. First, the
leading edge of a detached long-wavelength disturbance propagates at $v$,
independently of the source velocity. Second, for a rigid packet
$m(x,t)=m_0(x-X_0-ut)$ one has $\partial_tm=-u\,\partial_xm_0$, so the drive
is proportional to the source velocity and to the gradient of its profile; a
stationary pair in an adiabatically dressed fluid produces the quasistatic
deformation obtained from Eq.~\eqref{eq:appC_rhosol} with
$\partial_t\vartheta=0$, namely $\rho=-\chi ag_\rho m$. Switching on the
coupling at $t=0$ therefore emits fronts that propagate from the release region
along $x-X_0=\pm vt$, while the near-field component continues to follow
$X(t)$, which is the pattern seen in the compressible-regime column of
Fig.~\ref{fig:evolution}.

\subsection{Lattice equations for direct integration}

For the comparison with the many-body data it is preferable not to take the
continuum limit, since the source is only a few sites wide. Hamilton's
equations generated by the lattice Hamiltonian in Eq.~\eqref{eq:appA_lat} and
the lattice source in Eq.~\eqref{eq:appB_linear}, in units $a=1$ and using the
dimensionless time $\tau$, are
\begin{subequations}\label{eq:appC_lattice}
\begin{align}
  \partial_\tau\vartheta_i
  &=\Pi_i+\tilde g_\rho\,m_i(\tau),
  \label{eq:appC_lat_theta}\\
  \partial_\tau\Pi_i
  &=2r\nbar\,\big(\nabla^2\vartheta\big)_i
  -\tilde g_\theta\,m_i(\tau),
  \label{eq:appC_lat_Pi}
\end{align}
\end{subequations}
where
$(\nabla^2\vartheta)_i=\vartheta_{i+1}-2\vartheta_i+\vartheta_{i-1}$ is the
discrete Laplacian, not to be confused with the anisotropy $\Delta$ of
Eq.~\eqref{eq:HXXZ}. Equivalently, with the lattice current
\begin{equation}\label{eq:appC_lat_current}
  j_{i+1/2}=-2r\nbar\big(\vartheta_{i+1}-\vartheta_i\big),
\end{equation}
Eq.~\eqref{eq:appC_lat_Pi} is the discrete continuity equation
$\partial_\tau\Pi_i+(j_{i+1/2}-j_{i-1/2})=-\tilde g_\theta m_i$. Setting
$m_i=0$ and Fourier transforming with $\nabla^2\to-4\sin^2(q/2)$ gives
\begin{equation}\label{eq:appC_disp}
  \tilde\omega_q^2=8r\nbar\sin^2\frac q2,
  \qquad
  \tilde\omega_q=2\tilde v\left|\sin\frac q2\right|,
  \qquad
  \tilde v=\sqrt{2r\nbar},
\end{equation}
and the resulting dispersion is acoustic as $q\to0$ and bends below the linear law at larger $q$, so that the short-wavelength components of a narrow source lag behind the leading edge and account for part of the thickness of the numerically observed fronts.

One suitable scheme for integrating Eq.~\eqref{eq:appC_lattice} is the following. The source $m_i(\tau)$ is the magnon density measured in the many-body simulation, linearly interpolated in time between stored frames, so that the position, width, and velocity of the classical source coincide with those of the quantum one and no separate model of the trajectory is introduced. The fields $(\vartheta_i,\Pi_i)$ are propagated with a fourth-order Runge--Kutta integrator with an internal time step $\delta\tau_{\mathrm{RK}}$ small compared with $\tilde\omega_{q_{\max}}^{-1}$, chosen so that the output
times coincide exactly with the frames of the reference data. Open boundaries are implemented as discrete Neumann conditions,
$(\nabla^2\vartheta)_1=\vartheta_2-\vartheta_1$ and
$(\nabla^2\vartheta)_L=\vartheta_{L-1}-\vartheta_L$, which impose vanishing current at the ends and reproduce the reflecting boundaries of the finite chain; the analysis window is in any case restricted to times before reflected fronts return.

Two choices of initial data isolate different physical questions. Setting $\vartheta_i=\Pi_i=0$ describes a fluid that starts undeformed, matching the
quench protocol of Section~\ref{sec:model}, and produces both the attached deformation and the emitted fronts. Setting instead
$\Pi_i(0)=-\tilde g_\rho m_i(0)$, the quasistatic solution of
Eq.~\eqref{eq:appC_rhosol}, starts from an adiabatically dressed source and suppresses the part of the response that is a transient of the switch-on. The comparison of the two isolates the sound emitted by the motion of the source from the sound emitted by the switching itself. For the quench initial condition, $\Pi_i(\tau)$ is the predicted counterpart of the measured $\delta n_i(\tau)=\avg{\hat n_i(\tau)}-\avg{\hat n_i(0)}$, since both are density fluctuations relative to the same undeformed background.

%% file: sec_v3/appendix_two_magnon.tex
\section{Two-magnon bound state and trap-and-kick protocol}
\label{app:two_magnon}

This appendix derives the bound-state profile, dispersion, and internal size
given in Section~\ref{sec:model}, together with the anisotropy-matching
condition used in the preparation protocol. We work relative to the fully
down-polarized vacuum $\ket{\Downarrow}$ and keep the conventions of
Eq.~\eqref{eq:HXXZ}. The analytical derivation assumes an infinite chain, or
equivalently the bulk of a sufficiently long open chain.

\subsection{Bound-state solution}

A single up spin changes the field energy by $Jh_z$ and breaks two bonds,
changing the total Ising energy by $-J\Delta$. It hops by one site with
amplitude $J/2$. Two well-separated magnons therefore have the diagonal
excitation energy
\begin{equation}\label{eq:app_sep}
  \epsilon_{\mathrm{sep}}=2J(h_z-\Delta).
\end{equation}
When the two magnons occupy neighboring sites, the bond between them is
restored to its vacuum value and only the two outer bonds are affected, so
\begin{equation}\label{eq:app_adj}
  \epsilon_{\mathrm{adj}}=J(2h_z-\Delta).
\end{equation}
The relative nearest-neighbor interaction is the difference
\begin{equation}\label{eq:app_int}
  \epsilon_{\mathrm{adj}}-\epsilon_{\mathrm{sep}}=J\Delta,
\end{equation}
which is attractive for $\Delta<0$. This attraction is the only finite-range
interaction in the two-magnon sector of Eq.~\eqref{eq:HXXZ}, apart from
the hard-core constraint forbidding $x_1=x_2$, which is automatic for
spin-$1/2$. At fixed total momentum $K$, we write
\begin{equation}\label{eq:app_ansatz}
  \ket{\Psi_K}
  =\sum_{x_1<x_2}
  e^{\ii K(x_1+x_2)/2}\,
  \phi_K(\ell)\ket{x_1,x_2},
  \qquad
  \ell=x_2-x_1 .
\end{equation}
Projecting $\HXXZ$ onto this sector gives the relative-coordinate equations
\begin{subequations}\label{eq:app_rel}
\begin{align}
  \big(E-\epsilon_{\mathrm{sep}}\big)\phi_K(\ell)
  &=J\cos\frac K2
  \big[\phi_K(\ell-1)+\phi_K(\ell+1)\big],
  &&\ell\ge2,
  \label{eq:app_rel_bulk}\\
  \big(E-\epsilon_{\mathrm{adj}}\big)\phi_K(1)
  &=J\cos\frac K2\,\phi_K(2).
  \label{eq:app_rel_edge}
\end{align}
\end{subequations}
The absence of an $\ell=0$ amplitude implements the hard-core constraint, and
Eq.~\eqref{eq:app_rel_edge} differs from the bulk equation both by the missing
inward hopping and by the interaction energy in Eq.~\eqref{eq:app_int}.
Substituting a decaying solution
\begin{equation}\label{eq:app_decay}
  \phi_K(\ell)=\mathcal N_K A_K^{\ell-1},
  \qquad |A_K|<1,
\end{equation}
into Eq.~\eqref{eq:app_rel_bulk} gives
\begin{equation}\label{eq:app_bulkrel}
  E-\epsilon_{\mathrm{sep}}
  =J\cos\frac K2\left(A_K+A_K^{-1}\right),
\end{equation}
while Eq.~\eqref{eq:app_rel_edge} gives
\begin{equation}\label{eq:app_edgerel}
  E-\epsilon_{\mathrm{adj}}=J\cos\frac K2\,A_K .
\end{equation}
Combining the two equations with Eq.~\eqref{eq:app_int} eliminates $E$ and
fixes the decay constant,
\begin{equation}\label{eq:app_AK}
  A_K=\frac{\cos(K/2)}{\Delta},
\end{equation}
after which Eq.~\eqref{eq:app_edgerel} yields the bound-state band
\begin{equation}\label{eq:app_band}
  E_{\mathrm{B}}(K)
  =J(2h_z-\Delta)+\frac J\Delta\cos^2\frac K2 .
\end{equation}
The additive constant depends on the field and anisotropy conventions and has
no dynamical effect; the $K$-dependent term is the dispersion used in the main
text. The state is normalizable when $|\cos(K/2)|<|\Delta|$, which, for
$|\Delta|>1$, holds at every momentum but, for $|\Delta|<1$, only in a window
around $K=\pi$. On an infinite chain, the normalized profile and the mean
separation are
\begin{equation}\label{eq:app_size}
  \phi_K(\ell)=\sqrt{1-|A_K|^2}\,A_K^{\ell-1},
  \qquad
  \avg{\ell}_K=\frac{1}{1-|A_K|^2}.
\end{equation}
Differentiating Eq.~\eqref{eq:app_band} gives
\begin{equation}\label{eq:app_vel}
  v_{\mathrm{B}}(K)=-\frac{J\sin K}{2\Delta},
  \qquad
  E_{\mathrm{B}}''(K)=-\frac{J\cos K}{2\Delta}.
\end{equation}
For $\Delta<0$ and $0<K<\pi$, the group velocity is positive. At
$K_0=\pi/2$, the curvature vanishes, so the leading dispersive spreading of a
narrow momentum packet is suppressed and the residual broadening observed in
Section~\ref{sec:num} can be attributed to higher-order free dispersion and
coupling to the fluid rather than to the leading quadratic dispersion.

\subsection{Trapped-state preparation and momentum kick}

The localized initial state is the ground state of
\begin{equation}\label{eq:app_prep}
  \hat H_{\mathrm{prep}}
  =\HXXZ(\Delta_{\mathrm{prep}})
  +c_{\mathrm{trap}}\sum_i(i-X_0)^2\,\nm_i ,
\end{equation}
in the sector $\sum_i\nm_i=2$. For a sufficiently weak and smooth trap, the
center of mass is confined near $X_0$ with a Gaussian envelope, while the
relative coordinate retains the $K=0$ profile of Eq.~\eqref{eq:app_size},
that is,
$A(0,\Delta_{\mathrm{prep}})^{\ell-1}
=\Delta_{\mathrm{prep}}^{-(\ell-1)}$. Applying the kick of
Eq.~\eqref{eq:kick},
\begin{equation}\label{eq:app_kick}
  \hat U_{\mathrm{kick}}(k_0)\ket{x_1,x_2}
  =e^{\ii k_0(x_1+x_2)}\ket{x_1,x_2}
  =e^{\ii(2k_0)X}\ket{x_1,x_2},
\end{equation}
imprints the center-of-mass momentum $K_0=2k_0$ and leaves the relative
coordinate $\ell$ unchanged because the operator is diagonal in the magnon
positions. Requiring the prepared relative profile to match the profile
selected by Eq.~\eqref{eq:app_AK} for the evolution anisotropy yields
\begin{equation}\label{eq:app_match}
  \left|\frac{1}{\Delta_{\mathrm{prep}}}\right|
  =\left|\frac{\cos(K_0/2)}{\Delta_{\mathrm{evol}}}\right|,
\end{equation}
which, for the negative anisotropies used here and $0<K_0<\pi$, reduces to
Eq.~\eqref{eq:matching}. Without this matching, the pair would undergo an
internal quench at $t=0$ and would breathe in the relative coordinate,
thereby reshaping the one-body magnon density and contaminating the spatial-width
diagnostic of Fig.~\ref{fig:backreaction_size} with an
effect unrelated to the fluid response.

%% file: sec_v3/appendix_numerics.tex
\section{Tensor-network protocol and numerical control}
\label{app:num}

To complement the many-body results of Section~\ref{sec:num}, this appendix
describes the tensor-network representation of the coupled spin-boson system,
the preparation and real-time evolution protocol, and the numerical controls
used in the calculations.

\subsection{Composite local space}

Each matrix-product-state site combines a spin-$1/2$ degree of freedom with a
bosonic Fock space truncated to a maximum occupation $n_{\max}$,
\begin{equation}\label{eq:appE_local}
  \mathcal H_i
  =\mathbb C^2\otimes
  \operatorname{span}\{\ket{0},\ldots,\ket{n_{\max}}\},
  \qquad
  d_{\mathrm{loc}}=2(n_{\max}+1).
\end{equation}
The calculations use $n_{\max}=3$, corresponding to a local dimension
$d_{\mathrm{loc}}=8$. The basis ordering is
$\ket{\sigma,n}\mapsto\sigma(n_{\max}+1)+n+1$, where $\sigma=0,1$ labels the
two spin states. This choice keeps every term of Eq.~\eqref{eq:H} either
on-site or on a nearest-neighbor bond of the matrix-product-state ordering.
The exchange and Ising interactions, together with boson hopping, are bond
operators, whereas the displacement coupling $\HSB$ is strictly local.
Interleaving the spin and boson chains as separate sites would instead double
the chain length and place the two hopping terms at distance two.

The magnon number $\sum_i\nm_i$ is retained as a $U(1)$ quantum number and the
simulations are restricted to the two-magnon sector. The boson number cannot
be retained because $\HSB$ changes it by one. Consequently, the MPS must
represent superpositions of different boson-number sectors, which generally
requires larger bond dimensions than a number-conserving Bose--Hubbard
calculation.

\subsection{Preparation and evolution}

The protocol summarized in Fig.~\ref{fig:tn} proceeds in three stages.\\

\medskip
\noindent\emph{(a) Trapped composite source.}
The preparation Hamiltonian contains the XXZ chain with anisotropy
$\Delta_{\mathrm{prep}}$, the isolated Bose--Hubbard chain at the chosen $(r,s)$, and a weak harmonic trap acting on the two magnons, while the spin-boson coupling is set to zero. Its ground state is obtained in a single finite-system DMRG calculation in the sector $\sum_i\nm_i=2$. Since the two subsystems are decoupled during preparation, this state is equivalent to the product of the trapped two-magnon ground state of Eq.~\eqref{eq:app_prep} and
the Bose--Hubbard ground state of Eq.~\eqref{eq:HBH}. The resulting bosonic density profile defines $\avg{\hat n_i(0)}$ in
Eq.~\eqref{eq:density_response}.\\

The calculations use $J/U_B=1$, $h_z=0$, and $\Delta_{\mathrm{evol}}=-1.5$. The preparation anisotropy is fixed by
Eq.~\eqref{eq:app_match}, giving $\Delta_{\mathrm{prep}}=\sqrt{2}\Delta_{\mathrm{evol}}\simeq-2.1213$ for the momentum used here. The trap is centered at $X_0=L/2=40$ and has curvature $c_{\mathrm{trap}}/U_B=0.02$. The DMRG calculation is initialized from a product state with one boson per site and two adjacent magnons at $X_0$ and $X_0+1$; this state serves only as the initial seed for the variational optimization.\\

\medskip
\noindent\emph{(b) Release and momentum kick.}
The trap is removed and the kick
$\hat U_{\mathrm{kick}}(k_0)=\prod_j\exp(\ii k_0j\,\nm_j)$ of
Eq.~\eqref{eq:kick} is applied. Since it is diagonal in the magnon occupation,
the kick is a product of on-site phase gates and can be applied exactly to the
MPS without truncation. We use $k_0=\pi/4$, which imprints the total
center-of-mass momentum $K_0=2k_0=\pi/2$ while leaving the relative coordinate
and the bosonic state unchanged.\\

\medskip
\noindent\emph{(c) Real-time evolution.}
At $\tau=0$, the spin-boson interaction is switched on and the complete
Hamiltonian in Eq.~\eqref{eq:H} is evolved using the two-site TDVP~\cite{Haegeman2011,Haegeman2016,Paeckel2019}, as implemented in the
ITensor framework~\cite{Fishman2022}. This is a global coupling quench, but the
resulting drive is spatially localized because the polarized spin background
has $\nm_i=0$ and only the region occupied by the bound pair drives the
bosonic medium. The simulations use open chains of $L=80$ and are evolved up
to $\tau=30$, before boundary-reflected fronts return to the central analysis
window.

\subsection{Numerical control}

The DMRG preparation uses $120$ sweeps and a truncation cutoff of $10^{-9}$.
The maximum bond dimension is increased from $10$ to $20$ and $40$ during the first three sweeps and is then fixed at $80$. Noise amplitudes $10^{-4}$, $10^{-5}$, $10^{-6}$, and $10^{-7}$ are applied during the first four sweeps, respectively, and are set to zero afterward. No early-stopping criterion is used, so all $120$ sweeps are performed. The real-time evolution uses two-site TDVP with one sweep per time step, $\delta\tau=0.1$, a truncation cutoff of $10^{-6}$, and a maximum bond dimension of $160$. The evolution to $\tau=30$ therefore consists of $300$
TDVP steps. Observables are stored every ten steps, corresponding to
$\Delta\tau_{\mathrm{obs}}=1$ and yielding $31$ frames including the initial state. This cadence sets the temporal resolution of the density maps, the front trajectories, and the finite-difference derivative entering Eq.~\eqref{eq:balance_residual}.

The quantities used in the main text are the magnon density $m_i(\tau)$, the bosonic density $\avg{\hat n_i(\tau)}$, and the induced response $\delta n_i(\tau)$ defined in Eq.~\eqref{eq:density_response}. The magnon center of mass is extracted from $m_i(\tau)$ and determines the velocity of the bound pair. For Fig.~\ref{fig:backreaction_size}, the packet boundaries are extracted from the connected component containing the maximum of $m_i(\tau)$ at a relative threshold of $5\%$. The thresholded width $W_{\mathrm p}$ is the distance between these boundaries, and its early and late values are averaged over the first and last $20\%$ of $1\leq\tau\leq29$, respectively. The trajectories of the detached fronts are obtained from the ridges of $\delta n_i(\tau)$ after masking the density deformation attached to the pair. The bosonic current, the exact source $\hat{\mathcal S}_i$, and the finite-difference density derivative are retained for the continuity test in Fig.~\ref{fig:continuity}. Since the derivative is evaluated between consecutive stored frames, the continuity residual is defined only from the second frame onward.

These calculations use fixed production values of $n_{\max}=3$ and of the MPS cutoffs and bond dimensions specified above, rather than an automatic extrapolation in each numerical control. The conclusions therefore emphasize the propagation, broadening, and velocities of the resolved structures, whereas their amplitudes at the largest spin-boson couplings are more sensitive to the bosonic cutoff and MPS truncation.